\documentclass{ws-ijmpa}
\usepackage{epsf}
\newcommand\ap[3]{{{\it Ann. Phys. (NY) }{\bf #1},  #3 (#2)}}

\newcommand\eurc[3]{{{\it Eur. Phys. J. }{\bf C #1}, #3 (#2)}}

\newcommand\npb[3]{{{\it Nucl. Phys. }{\bf B #1},  #3 (#2)}}
\newcommand\npbp[3]{{{\it Nucl. Phys. }{\bf B #1}, {\it(Proc. Suppl.)} #3 (#2)}}
\newcommand\plb[3]{{{\it Phys. Lett. }{\bf B #1}, #3 (#2)}}
\newcommand\prd[3]{{{\it Phys. Rev. }{\bf D #1}, #3  (#2)}}
\newcommand\pr[3]{{{\it Phys. Rev. }{\bf #1}, #3 (#2)}}

\newcommand\rmp[3]{{{\it Rev. Mod. Phys. }{\bf #1}, #3 (#2)}}
\newcommand\zpc[3]{{{\it Z. Physik }{\bf C #1}, #3 (#2)}}

\newcommand\nc[3]{{{\it Nuovo Cim. }{\bf #1}, #3 (#2) }}

\newcommand\ibid[3]{    {\it ibid. }{\bf #1}, #3 (#2) }
\newcommand\ijmpa[3]{{{\it Int. J. Mod. Phys. }{\bf A #1}, #3 (#2)}}

\newcommand\hepph[1]{\tt hep-ph/#1}

\def\eq#1{{Eq.~(\ref{#1})}}
\def\eqs#1#2{{Eqs.~(\ref{#1})--(\ref{#2})}}

\let\vev\VEV

\def\Im{\mathop{\mbox{Im}}}
\def\Re{\mathop{\mbox{Re}}}
\def\Tr{\mathop{\mbox{Tr}}\,}
\def\etal{{\it et\,al.}}

\def\De{\Delta}
\def\g{\gamma_{5}}
\newcommand{\bea}{\begin{eqnarray}}
\newcommand{\beq}{\begin{equation}}
\newcommand{\eea}{\end{eqnarray}}
\newcommand{\eeq}{\end{equation}}
\newcommand{\nnu}{\nonumber}
\newcommand{\m}{\multicolumn}
\begin{document}
\markboth{E. I. Lashin}
{CP conserved nonleptonic $K \rightarrow \pi \pi \pi $ decays in the Chiral
Quark Model.}
\title{CP CONSERVED NONLEPTONIC $K \rightarrow \pi \pi \pi $ DECAYS
IN THE CHIRAL QUARK MODEL.}
\author{\footnotesize E. I. LASHIN}
\address{The Abdus Salam ICTP, P. O. Box 586, 34100 Trieste, Italy\\
Department of Physics and Astronomy, College of Science,\\
King Saud University, P. O. Box 2455, 11451  Riyadh, Saudi Arabia\\
Department of Physics, Faculty of Science, Ain Shams
University, Cairo, Egypt\\elashin@ictp.it }

\maketitle

\begin{history}
\received{26 1 2005}
\revised{8 7 2005}
\end{history}

\begin{abstract}
The CP conserved non leptonic $K\rightarrow \pi \pi \pi$ decays
are discussed within the chiral quark model, including chiral
perturbation theory corrections. For the chiral loop correction,
they are presented in a way to be more manageable and easier for use.
Furthermore a new identity for the imaginary part of the chiral loop
corrections is derived.   All amplitudes are parameterized
in terms of quark and gluon condensate and the constituent quark
mass.  The same values for these parameters that can be obtained
by a fit  of the $\Delta I  = 1/2$ rule in $K\rightarrow 2 \pi$
give a reasonably good fit in the $K\rightarrow 3 \pi$ case. We
compare with the work of other groups.

\end{abstract}

\ccode{PACS numbers: 13.20.Eb, 12.39.Fe, 14.40.Aq,11.30.Rd}

\section {Introduction}
Kaon physics provides us with many interesting phenomena, among them the
striking $\Delta I = 1/2$ selection rule in the $K\rightarrow \pi \pi \pi$
decay. The decay of a Kaon into three pions has a long history and many
attempts have been made to estimate its decay amplitudes. The first
estimates were done using current algebra methods or tree level
lagrangian, see Refs.~\refcite{corn}--\refcite{dev}. Afterwards, Chiral
Perturbation Theory (CHPT)~\cite{vein}\cdash \cite{chpt2} at tree level
was used in Refs.~\refcite{treech1}--\refcite{treech3}.

The first one-loop calculation of the $K\rightarrow \pi \pi \pi$ decay
amplitudes in (CHPT) was presented in Ref.~\refcite{meson}, but unfortunately
the details of this work were lost as mentioned in Ref.~\refcite{bij}. Recently,
the first full published result, to our knowledge,  appeared in
Ref.~\refcite{bij}, and was later confirmed in Ref.~\refcite{gam}. More recently,
this was followed by including the isospin breaking effects
(see Refs.~\refcite{iso1}--\refcite{iso4} and references therein).

The aim of the present work is to give an estimate for
$K\rightarrow \pi\pi\pi$ decay amplitudes based on the chiral
quark model ($\chi$QM) \cite{corn,vein}$^{,}$\cite{QMM1}\cdash\cite{QM1}
approach. The model has been
applied to the $K\rightarrow \pi \pi$ decays \cite{I}\cdash\cite{VI}
and the $ K^0 - \bar{K}^0$ mixing \cite{V,IV} leading to
satisfactory results. The strategy is to start with the quark
effective Lagrangian, which it turns out splits the physics
contributions into short- and long-distances  encoded in the
Wilson coefficients and hadronic matrix elements respectively.

In evaluating hadronic matrix elements we exploit the chiral quark
model as an effective link between QCD and chiral perturbation
theory, and thus the hadronic matrix elements can be estimated
using chiral perturbation techniques.  The long distance contributions
computed this way are eventually matched to the short-distance Wilson
coefficients and the results are compared with the experimental
values.

The chiral quark model furnishes us with three input parameters
namely the constituent quark mass M, the gluon condensate $\langle
\frac{\alpha_s}{\pi} G G \rangle$ and the quark  condensate
$\langle \overline{q} q \rangle$. A detailed discussion about
these parameters and their determination can be found in
Refs.~\refcite{I,II}.

In order for the present paper to be as self contained as
possible, we have included in the following first two sections
the relevant Lagrangian and a brief introduction to the $\chi$QM.
After then the paper is organized as follows:\\
in section~4 we present the lowest order chiral lagrangian, while
in section~5 the isospin decomposition of the decay amplitudes is
shown. Section~6 is devoted for computing the leading and
next-to-leading order for the decay amplitudes.  In section~7 we
discuss the relevant input parameters for the chiral quark model
and in section~8 we do the fit with the experimental data and
compare with the work of other groups. Finally section~9 is for
conclusion. We collect all useful formulas and relations relevant
for our work in the appendix.

\section{The Quark Effective Lagrangian}
The $\Delta S=1$
quark effective Lagrangian at a scale $\mu < m_c$ can be written
as~\cite{GW1}\cdash\cite{GW4}
 \beq
{\cal L}_{\Delta S = 1} = -
\frac{G_F}{\sqrt{2}} V_{ud}\,V^*_{us} \sum_i \Bigl[
z_i(\mu) + \tau y_i(\mu) \Bigr] Q_i (\mu)
 \, . \label{ham}
\label{qwh}
\eeq

The $Q_i$ are effective
four-quark operators obtained by integrating out, in the standard
model, the vector bosons and the heavy quarks $t,\,b$ and $c$.
A convenient and by now standard
basis includes the following ten quark operators:
 \beq
\begin{array}{rcl}
Q_{1} & = & \left( \overline{s}_{\alpha} u_{\beta}  \right)_{\rm V-A}
            \left( \overline{u}_{\beta}  d_{\alpha} \right)_{\rm V-A}
\, , \\
Q_{2} & = & \left( \overline{s} u \right)_{\rm V-A}
            \left( \overline{u} d \right)_{\rm V-A}
\, , \\
Q_{3,5} & = & \left( \overline{s} d \right)_{\rm V-A}
   \sum_{q} \left( \overline{q} q \right)_{\rm V\mp A}
\, , \\
Q_{4,6} & = & \left( \overline{s}_{\alpha} d_{\beta}  \right)_{\rm V-A}
   \sum_{q} ( \overline{q}_{\beta}  q_{\alpha} )_{\rm V\mp A}
\, , \\
Q_{7,9} & = & \frac{3}{2} \left( \overline{s} d \right)_{\rm V-A}
         \sum_{q} \hat{e}_q \left( \overline{q} q \right)_{\rm V\pm A}
\, , \\
Q_{8,10} & = & \frac{3}{2} \left( \overline{s}_{\alpha}
                                                 d_{\beta} \right)_{\rm V-A}
     \sum_{q} \hat{e}_q ( \overline{q}_{\beta}  q_{\alpha})_{\rm V\pm A}
\, ,
\end{array}
\label{Q1-10}
\eeq
where $\alpha$, $\beta$ denote color indices
($\alpha,\beta =1,\ldots,N_c$) and $\hat{e}_q$  are quark charges.
Color indices for the color singlet operators are omitted. The
subscripts $(V\pm A)$ refer to $\gamma_{\mu} (1 \pm \gamma_5)$
combinations. We recall that $Q_{1,2}$ stand for the $W$-induced
current--current operators, $Q_{3-6}$ for the QCD penguin
operators and $Q_{7-10}$ for the electroweak penguin (and box)
ones.

Even though not all the operators in \eq{Q1-10} are independent, this basis
is of particular interest for
the present numerical analysis because it is that employed
for the calculation of the Wilson coefficients
to the NLO in $\alpha_s$.~\cite{NLO1}\cdash\cite{NLO6}

In the present paper we will mainly discuss the features related
to the first six
operators in \eq{Q1-10} because the electroweak penguins
$Q_{7-10}$ have their contributions
suppressed by the smallness of their $CP$ conserving Wilson coefficients.

The functions $z_i(\mu)$ and $y_i(\mu)$ are the
 Wilson coefficients and $V_{ij}$ the
Koba\-ya\-shi-Mas\-kawa (KM) matrix elements; $\tau = - V_{td}
V_{ts}^{*}/V_{ud} V_{us}^{*}$. The numerical values of the Wilson
coefficients at a given scale depend on $\alpha_s$, for which we
use the  recent  average~\cite{alfa} $\alpha_s (m_Z) = 0.1189 \pm
0.002$, corresponding to $ \Lambda^{(4)}_{\rm QCD} = 340 \pm 40$.
We match the Wilson coefficients at the $m_W$ scale with the full
electroweak theory by using the LO $\overline{MS}$ running top
mass $m_t(m_W)=177\pm 7$ GeV which corresponds to the pole mass
$m_t^{pole}=175\pm 6$ GeV.~\cite{mt} For the remaining quark
thresholds we take $m_b(m_b) =4.4$ GeV and $m_c(m_c)=1.4$ GeV.

\section{The Chiral Quark Model}
In order to evaluate the bosonization
of the quark operators in \eq{Q1-10}
we exploit  the $\chi$QM approach
which provides an effective link between  QCD and
chiral perturbation theory.

The $\chi$QM  can be thought of as
the mean field approximation of the extended Nambu-Jona-Lasinio
(ENJL) model for low-energy
QCD. A detailed discussion of the ENJL model and
its relationship with QCD---as well as with the $\chi$QM---
can be found, for instance, in Refs.~\refcite{BBdeR1}--\refcite{BBdeR3}.

In the $\chi$QM, the light (constituent)
quarks are coupled to the Goldstone mesons by
the term
\beq
 {\cal{L}}^{\rm int}_{\chi \mbox{\scriptsize QM}} =
- M \left( \overline{q}_R \; \Sigma q_L +
\overline{q}_L \; \Sigma^{\dagger} q_R \right) \, ,
\label{M-lag}
\eeq
where $q^T\equiv (u,d,s)$ is the quark flavor triplet, and
the $3\times 3$ matrix
\beq
\Sigma \equiv \exp \left( \frac{2i}{f} \,\Pi (x)  \right)
\label{sigma}
\eeq
contains the pseudoscalar meson octet
\beq
\Pi (x) =
\frac{1}{\sqrt{2}}\,
\left[
\begin{array}{ccc}
\frac{1}{\sqrt{2}}\, \pi^0 (x) + \frac{1}{\sqrt{6}}\, \eta_8 (x) & \pi^+ (x) & K^+ (x) \\
\pi^- (x) & -\frac{1}{\sqrt{2}}\, \pi^0 (x) +  \frac{1}{\sqrt{6}}\, \eta_8 (x) & K^0 (x)\\
K^- (x) & \bar{K}^0 (x) & - \frac{2}{\sqrt{6}} \, \eta_8 (x)
\end{array}
\right] \, .
\label{defseg}
\eeq
The scale
$f$ is identified at the tree level with the  pion decay constant $f_\pi$.

The $\chi$QM has been discussed in several works over the
years.~\cite{QMM1}\cdash\cite{QM1} We opted for the somewhat more restrictive
definition suggested in Ref.~\refcite{QM1} (and there referred to as
the QCD effective action model) in which the meson degrees of
freedom do not propagate in the original Lagrangian.

The QCD gluonic fields are considered as
integrated out down to the  chiral breaking  scale $\Lambda_\chi$, here  acting
as an infrared cut-off. The effect of the
remaining low-frequency modes are assumed to be
well-represented by gluonic vacuum condensates, the leading contribution
coming from
\beq
\langle \frac{\alpha_s}{\pi} G G \rangle \, .
\eeq
The constituent quarks are taken to be
propagating in the fixed background of the soft gluons.
This defines an effective QCD Lagrangian
$ {\cal{L}}^{\rm eff}_{\rm QCD} (\Lambda_\chi) $,
whose propagating fields are the $u,d,s$ quarks.
The complete $\chi$QM Lagrangian is therefore given by
\beq
{\cal{L}}_{\chi \mbox{\scriptsize QM}} =
{\cal{L}}^{\rm eff}_{\rm QCD} (\Lambda_\chi) +
 {\cal{L}}^{\rm int}_{\chi \mbox{\scriptsize QM}}\ .
\label{LchiQM}
\eeq

The ${\cal{L}}_{\chi \mbox{\scriptsize QM}}$ interpolates between
the chiral breaking scale
$\Lambda_{\chi}$ and $M$ (the constituent quark mass).
The three light quarks ($u,\ d,\ s$) are the only dynamical degrees of freedom
present within this range.
The total quark masses are given by $M+m_q$ where
$m_q$ is the current quark mass in the QCD Lagrangian.
The Goldstone bosons and
the soft QCD gluons are taken in our approach as external fields.
A kinetic term for the mesons, as well as the complete
chiral Lagrangian,
is generated and determined by
integrating out the constituent quark degrees of freedom of the model.
The $\Delta S =1$  weak
chiral Lagrangian  thus becomes the effective theory of
the $\chi$QM
below the constituent quark mass scale $M$.
In the matching process, the many coefficients of the chiral
Lagrangian are determined---to the order $O(\alpha_s N_c)$ in our
computation---in
terms of $M$, the quark and gluon condensates.
We neglect heavier scalar, vector and axial meson multiplets.

In conventional chiral perturbation theory the scale dependence of
meson loops renormalization is canceled out, by construction,
through the $O(p^4)$ counterterms in the chiral Lagrangian. While
in our approach this is maintained for the strong sector of the
chiral Lagrangian. The tree-level counterterms of the weak sector
are taken to be $\mu$ independent and a scale dependence is
introduced in the hadronic matrix elements via the meson loops,
evaluated in dimensional regularization with minimal subtraction.
This scale dependence is eventually matched with that of the
Wilson coefficients. The stability of the final results against $\mu$
variation is numerically checked.

\section{Lowest order chiral Lagrangian}

At the leading $O(p^2)$  order in chiral expansion, the strong interaction
between the $SU(3)$ Goldstone bosons is described by the following
effective Lagrangian:
\beq
{\cal L}_{strong}^{(2)} = \frac{f^2}{4}
\Tr (D_\mu \Sigma^{\dagger} D^\mu \Sigma )
+  \frac{f^2}{2} B_0 \Tr ( {\cal M}  \Sigma^{\dagger} +
\Sigma {\cal M}^{\dagger} )
\label{sham}
\eeq
 where ${\cal M}$ is the mass matrix of the three light quarks
 $(u, d$ and $s)$ and $B_0$ is defined by $\langle \bar{q}_i q_j
 \rangle = - f^2 B_0 \delta_{ij}$. The $3 \times 3 $ matrix $\Sigma$ is
 the same as defined in \eqs{sigma}{defseg}.
The scale $f$ is identified with pion
decay constant $f_\pi$ in lowest order.

To the same order, the complete $\Delta S = 1$ weak chiral Lagrangian
in \eq{qwh} can be found in Ref.\cite{I}. Here we just quote the result
for the operators $Q_{1-6}$, which turn out to be relevant for the $CP$
conserving amplitude $K \rightarrow \pi \pi \pi$. The electroweak penguins
$Q_{7-10}$ give a negligible contribution due to the smallness of
their Wilson coefficients.

The
bosonization of the relevant operators leads to
\bea
{\cal L}^{(2)}_{\Delta S = 1}  & = &G_{\underline{8}} (Q_{3-6}) \Tr \left(
\lambda^3_2 D_\mu \Sigma^{\dag}
D^\mu \Sigma
\right) +  \nnu \\
 & & \: G_{LL}^a (Q_{1,2}) \,
\Tr \left(  \lambda^3_1 \Sigma^{\dag} D_\mu \Sigma \right)
\Tr \left( \lambda^1_2 \Sigma^{\dag} D^\mu  \Sigma \right) +  \nnu \\
 & & \: G_{LL}^b (Q_{1,2})\, \Tr \left( \lambda^3_2 \Sigma^{\dag} D_\mu
\Sigma \right)
\Tr \left(  \lambda^1_1 \Sigma^{\dag} D^\mu \Sigma \right)
\label{chi-lag} \, ,
\eea
where $\lambda^i_j$ are combinations of Gell-Mann
$SU(3)$ matrices defined by $(\lambda^i_j)_{lk} = \delta_{il}\delta_{jk}$
and $\Sigma$ is defined in \eq{sigma}.
The covariant
derivatives in \eq{chi-lag} are taken with respect to the external
gauge fields whenever they are present.

\begin{table}[h]
\tbl{Values of the relevant $\Delta S = 1$
weak chiral coefficients for two different
regularization schemes: HV and NDR. The inclusion of the Wilson
coefficients of the effective quark operators $Q_i$ is
understood.}
{\begin{tabular}{@{}|l|l|@{}}
\hline
\hspace*{2.5cm} HV & \hspace{2.5cm} NDR
\\
\hline
$ G_{LL}^a(Q_1)  =  - \frac{1}{N_c} f_\pi^4  \left( 1 - \delta_{\vev{GG}}
\right) \nnu $  & $ G_{LL}^a(Q_1) =  - \frac{1}{N_c} f_\pi^4  \left( 1 - \delta_{\vev{GG}}
\right) \nnu $
\\
$ G_{LL}^a(Q_2) =  - f_\pi^4  \nnu $  & $ G_{LL}^a(Q_2)  =  - f_\pi^4 $ \nnu
\\
$ G_{LL}^b(Q_1)  =  - f_\pi^4  \nnu $  & $ G_{LL}^b(Q_1) =  - f_\pi^4  \nnu $
\\
$ G_{LL}^b(Q_2)  = - \frac{1}{N_c} f_\pi^4 \left( 1 - \delta_{\vev{GG}} \right)
  \nnu $ &
$ G_{LL}^b(Q_2)  = - \frac{1}{N_c} f_\pi^4 \left( 1 - \delta_{\vev{GG}} \right)
  \nnu $
\\
 $ G_{\underline{8}} (Q_3)  =    f_\pi^4 \frac{1}{N_c} \left( 1 -
\delta_{\vev{GG}}
\right) \nnu $ &  $ G_{\underline{8}} (Q_3)  =   f_\pi^4 \frac{1}{N_c} \left( 1 -
\delta_{\vev{GG}}
- 6 \frac{M^2}{\Lambda_\chi^2}
\right) \nnu $
\\
$ G_{\underline{8}} (Q_4)  =    f_\pi^4 \nnu $ & $G_{\underline{8}} (Q_4)  =   f_\pi^4 \left( 1 - 6 \frac{M^2}{\Lambda_\chi^2}
\right) \nnu $
\\
$ G_{\underline{8}} (Q_5)  =   \frac{2}{N_c} \,
\frac{\vev{\bar{q}q}}{M} f_\pi^2 \, \left( 1 - 6\,
\frac{M^2}{\Lambda_{\chi}^2} \right) \nnu $  & $ G_{\underline{8}} (Q_5) =  \frac{2}{N_c} \,
\frac{\vev{\bar{q}q}}{M} f_\pi^2 \, \left( 1 - 9\,
\frac{M^2}{\Lambda_{\chi}^2} \right)\nnu $
\\
 $ G_{\underline{8}} (Q_6)  =  2 \,
\frac{\vev{\bar{q}q}}{M} f_\pi^2 \, \left( 1 - 6\,
\frac{M^2}{\Lambda_{\chi}^2} \right)  \nnu $ & $ G_{\underline{8}} (Q_6) =  2 \,
\frac{\vev{\bar{q}q}}{M} f_\pi^2 \, \left( 1 - 9\,
\frac{M^2}{\Lambda_{\chi}^2} \right)  \nnu $
\\
\hline
\end{tabular}
\label{Gchi}
}
\end{table}

The notation for the chiral coefficients\footnote{It should be
understood that the chiral coefficients are multiplied by the
appropriate factors containing Fermi constant and Wilson
coefficients, but for convenience these factors are not shown.}
$G_{\underline{8}} (Q_{3-6})$, $G_{LL}^a(Q_{1,2})$ and $G_{LL}^b
(Q_{1,2})$ reminds us of their chiral properties:
$G_{\underline{8}}$ represents the $(\underline{8}_L \times
\underline{1}_R)$ part of the interaction induced in QCD by the
gluonic penguins, while the two terms proportional to $G_{LL}^a$
and $G_{LL}^b$ are admixture of the $(\underline{27}_L \times
\underline{1}_R)$ and the $(\underline{8}_L \times
\underline{1}_R)$ part of the interaction, induced by left-handed
current-current operators. These coefficients have been evaluated
in two different schemes of regularization HV and NDR, and the
results are given in Table~(\ref{Gchi}). The parameter
$\delta_{\vev{GG}}$ is defined as follows,
\beq
\delta_{\vev{GG}}={N_c\over 2}{\vev{\alpha_s\,GG\over \pi}\over 16\,
\pi^2\,f_\pi^2},
\label{ggc}
\eeq
where $N_c$ is the number of colors. The chiral symmetry breaking scale
is identified with $\Lambda_\chi=2\,\pi\,\sqrt{6\over N_c}\,f_\pi$.

In Table~(\ref{wilson}) we report the Wilson
coefficients of the first six operators at the scales $\mu = 0.8$,
$\mu = 1 $ GeV in the NDR and HV $\gamma_5$-schemes, respectively.
Since $\Re \tau$  in \eq{ham} is of $O(10^{-3})$, the CP
conserving part of $K \rightarrow \pi \pi \pi $ transition is
controlled by the  coefficients $z_i$, which do not depend on
$m_t$.
\begin{table}[h]
\tbl{NLO Wilson coefficients at $\mu=0.8$ GeV in the NDR and in the HV
scheme ($\alpha=1/128$).
The corresponding values at $\mu=m_W$ are given in parenthesis.
In addition one has $z_{3-6}(m_c)=0$. The coefficients
$z_i(\mu)$ do not depend on $m_t$.}
{\begin{tabular}{@{}|c|r r||r r||r r|@{}}
\hline
$\Lambda_{QCD}^{(4)}$ & \multicolumn{2}{c||}{ 300 MeV }
                      & \multicolumn{2}{c||}{ 340 MeV }
                      & \multicolumn{2}{c| }{ 380 MeV } \\
\hline
$\alpha_s(m_Z)_{\overline{MS}}$
                      & \multicolumn{2}{c||}{ 0.113 }
                      & \multicolumn{2}{c||}{ 0.119 }
                      & \multicolumn{2}{c| }{ 0.125 } \\
\hline \hline
\multicolumn{7}{|c|}{NDR}\\
\hline
$z_1$&$(0.0503)$&$-0.591$&$(0.0533)$&$-0.649$&$(0.0557)$&$-0.707$ \\
\hline
$z_2$&$(0.982)$&$1.336$&$(0.981)$&$1.377$&$(0.980)$&$1.421$ \\
\hline
$z_3$&$$&$0.0250$&$  $&$0.0333$&$ $&$0.0457$ \\
\hline
$z_4$&$$&$-0.0625$&$  $&$-0.0799$&$ $&$-0.1045$ \\
\hline
$z_5$&$$&$0.0088$&$  $&$0.0081$&$ $&$0.0054$ \\
\hline
$z_6$&$$&$-0.0672$&$  $&$-0.0881$&$ $&$-0.1193$ \\
\hline
\hline
\multicolumn{7}{|c|}{HV}\\
\hline
$z_1$&$(0.0320)$&$-0.769$&$(0.0339)$&$-0.879$&$(0.0355)$&$-1.016$ \\
\hline
$z_2$&$(0.988)$&$1.468$&$(0.987)$&$1.554$&$(0.987)$&$1.666$ \\
\hline
$z_3$&$$&$0.0190$&$  $&$0.0276$&$ $&$0.0392$ \\
\hline
$z_4$&$$&$-0.0392$&$  $&$-0.0505$&$ $&$-0.0664$ \\
\hline
$z_5$&$$&$0.0084$&$  $&$0.0097$&$ $&$0.0109$ \\
\hline
$z_6$&$$&$-0.0372$&$  $&$-0.0481$&$ $&$-0.0637$ \\
\hline
\end{tabular}
\label{wilson}
}
\end{table}

In comparing with other works,\cite{meson,bij,DI} it is
convenient to write the $\Delta S =1$ weak chiral Lagrangian in
the following form \bea {\cal L}^{(2)}_{\Delta S = 1} & = &
g_{\underline{8}} \Tr \left( \lambda^3_2 D_\mu \Sigma^{\dag}
D^\mu \Sigma\right) \ \nnu \\
&   & +
g_{\underline{27}} \left[
\Tr \left( \lambda^3_2 \Sigma^{\dag} D_\mu \Sigma \right)
\Tr \left( \lambda^1_1 \Sigma^{\dag} D^\mu  \Sigma \right) +
\frac{2}{3} \Tr \left( \lambda^3_1 \Sigma^{\dag} D_\mu
\Sigma \right)
\Tr \left(  \lambda^1_2 \Sigma^{\dag} D^\mu \Sigma \right) \right] \ , \nnu \\
\label{chiwyler} \, \eea which involves two couplings $g_8$ and
$g_{27}$ representing respectively the octet $ (8_L, 1_R)$ and the
twenty--seven $(27_L,1_R)$ couplings. There is only one convention
used for the octet part, however there are different ones for the
twenty--seven part, so attention should be paid in comparing
results.

In the $\chi$QM,
these couplings can be written as
\bea
g_{\underline{8}} & = &  \frac{G_F}{\sqrt{2}} V_{ud} V^*_{us} \frac{1}{5} \left\{
- 3 G^a_{LL} ( Q_1 ) z_1(\mu) -  3 G^a_{LL} ( Q_2 ) z_2(\mu) +
 2 G^b_{LL} ( Q_1 ) z_1(\mu) + \right. \ \nnu \\
& &  \left. 2 G^b_{LL} ( Q_2 ) z_2(\mu) +
5 \left[G_{\underline{8}} (Q_3) z_3(\mu) + G_{\underline{8}} (Q_4) z_4(\mu) +
 G_{\underline{8}} (Q_5) z_5(\mu) +   G_{\underline{8}} (Q_6) z_6(\mu) \right] \right\}
\ , \nnu \\
g_{\underline{27}} & = &  \frac{G_F}{\sqrt{2}} V_{ud} V^*_{us} \frac{3}{5} \left\{
\left( G^a_{LL} ( Q_1 ) + G^b_{LL} ( Q_1 ) \right) z_1(\mu) +
\left( G^a_{LL} ( Q_2 ) + G^b_{LL} ( Q_2 ) \right) z_2(\mu) \right\} ,
\label{ourcoupling}
\eea
where $ G_a$, $G_b$ and $G_8$ are the chiral coefficients displayed in
Table~(\ref{Gchi}), while $z_i(\mu)$ are the Wilson coefficients given in
Table~(\ref{wilson}).
The coupling $g_{\underline{8}} $ and $g_{\underline{27}}$ depend
implicitly on the parameters $\langle \frac{\alpha_s}{\pi} G G \rangle$,
$\langle \overline{q} q \rangle$ and $M$.

It seems redundant to express $g_{\underline{8}} $ and
$g_{\underline{27}}$ in terms of three input parameters, but one
should keep in mind that the complete $ \Delta S = 1$ weak chiral
Lagrangian  contains additional  terms transforming as $ (8_L ,
8_R ) $ with their associated couplings. Such terms plays an
important role in the study of CP violation which is not the
subject of our present work, a detailed discussion for these terms
can be found in Ref.~\refcite{I}.

\section{$K\rightarrow \pi \pi \pi $ Formalism}
There are four distinct channels for $K\rightarrow \pi \pi \pi$
decays:
\beq
\begin{array}{llll}
K^+ & \rightarrow & \pi^+ \pi^+ \pi^-, & \hspace{2cm} (I = 1, 2), \nnu\\
K^+ & \rightarrow & \pi^0 \pi^0 \pi^+,  & \hspace{2cm} (I = 1, 2), \nnu\\
K^0  &\rightarrow & \pi^+ \pi^- \pi^0,   & \hspace{2cm} (I = 0,1, 2), \nnu \\
K^0  &\rightarrow & \pi^0 \pi^0 \pi^0,   & \hspace{2cm} (I = 1).\nnu\\
\label{amp}
\end{array}
\eeq
 Near each channel we have indicated the final state isospin
assuming $\Delta I \leq \frac{3}{2}$.

In order to write the transition amplitude, it is convenient to introduce the following kinematical
 variables :
\beq
s_i = (p_K - p_i)^2 \hspace{1cm} \mbox{and} \hspace{1cm} s_0
= \frac{1}{3} (s_1 + s_2 + s_3) = \frac{1}{3} m_K^2 + m_\pi^2,
\eeq
where $p_K$ and $p_i$ denote Kaon and $\pi_i$ momenta
($\pi_3$ indicates the odd pion i.e the third
 pion in each of the final
states $  \pi^+ \pi^+ \pi^-, \pi^0 \pi^0 \pi^+ , \pi^+ \pi^- \pi^0 $ ).

We define the dimensionless Dalitz plot variables.
\beq Y
=\frac{s_3 - s_0}{m_\pi^2}\hspace{2cm} \mbox{and} \hspace{2cm} X =
\frac{s_2 - s_1}{m_\pi^2} . \label{defxy} \eeq
Expanding the four
amplitudes in \eq{amp} in powers of $X$ and $Y$ up to quadratic
terms, we get
\bea A_{++-} = A(K^+ \rightarrow \pi^+ \pi^+ \pi^-)
& = &
2 a_c + (b_c + b_2) Y \nnu \\
& & + 2 c_c ( Y^2 + X^2/3 ) + (d_c + d_2) ( Y^2 - X^2/3 ), \nnu \\
A_{00+} = A(K^+ \rightarrow \pi^0 \pi^0 \pi^+) & = & a_c - (b_c - b_2) Y \nnu \\
& & + c_c ( Y^2 + X^2/3 ) - (d_c - d_2 )  ( Y^2 - X^2/3 ) , \nnu \\
A_{+-0} = \sqrt{2} A (K^0 \rightarrow \pi^+ \pi^- \pi^0) & = &
a_n - b_n Y - \frac{2}{3} b_2 X \nnu \\
& & + c_n  ( Y^2 + X^2/3 ) - d_n ( Y^2 - X^2/3 ) + \frac{4}{3} d_2 X Y, \nnu \\
A_{000} = \sqrt{2} A (K^0 \rightarrow \pi^0 \pi^0 \pi^0) & = & 3
\left[ a_n + c_n  ( Y^2 + X^2/3 ) \right].
\label{rep1}
\eea
Here, we follow the notations and conventions of
Refs.~\refcite{DI}--\refcite{MP}.

A representation alternative  to that in \eq{rep1}, which has been
adopted in fits to experimental data,~\cite{meson,dev} is the
expansion in terms of amplitudes with definite isospin selection
rules. Such an expansion can be written as
\bea
A(K_L \rightarrow \pi^+ \pi^- \pi^0 ) & = & (\alpha_1 + \alpha_3) - (\beta_1 + \beta_3) Y \nnu \\
& & + ( \zeta_1 - 2 \zeta_3 ) ( Y^2 + X^2/3 ) + (\xi_1 - 2 \xi_3 ) ( Y^2 - X^2/3 ), \nnu \\
A(K_L \rightarrow \pi^0 \pi^0 \pi^0 ) & = & 3 (\alpha_1 + \alpha_3)
         + 3  ( \zeta_1 - 2 \zeta_3 ) ( Y^2 + X^2/3 ) , \nnu  \\
 A(K^+ \rightarrow \pi^+ \pi^+ \pi^-) & = &
- (2 \alpha_1 - \alpha_3) - (\beta_1 - \frac{1}{2}  \beta_3 + \sqrt{3} \gamma_3 ) Y \nnu \\
& & - 2 (\zeta_1 + \zeta_3 ) ( Y^2 + X^2/3 ) + (\xi_1 + \xi_3 - \xi_3^\prime) ( Y^2 - X^2/3), \nnu \\
A(K^+ \rightarrow \pi^0 \pi^0 \pi^+) & = &
 - \frac{1}{2} (2 \alpha_1 - \alpha_3) + (\beta_1 - \frac{1}{2}  \beta_3 - \sqrt{3}\, \gamma_3) Y \nnu \\
& & - (\zeta_1 + \zeta_3 ) ( Y^2 + X^2/3 ) -  (\xi_1 + \xi_3 + \xi_3^\prime) ( Y^2 - X^2/3 ) , \nnu \\
A (K_S \rightarrow \pi^+ \pi^- \pi^0) & = &
\frac{2}{3}\, \sqrt{3}\, \gamma_3 X - \frac{4}{3} \xi_3^\prime X Y .
\label{rep2}
\eea
In \eq{rep2}, the subscripts $1$ and $3$ refer to $\Delta I = 1/2 $ and
$\Delta I = 3/2 $, respectively. The relation between the amplitudes in \eq{rep1} and those
in \eq{rep2} is easily found to be
\beq
\begin{array}{lllllll}
a_c &=& - \alpha_1 + \alpha_3/2,  & &   a_n &=& \alpha_1 + \alpha_3, \\
b_c &=& - \beta_1 + \beta_3/2 ,  & &     b_n &=& \beta_1 + \beta_3,  \\
c_c &=& - \zeta_1 - \zeta_3,       & &     c_n &=& \zeta_1 - 2 \zeta_3, \\
d_c &=& \xi_1 + \xi_3,           & &     d_n &=& - \xi_1 + 2 \xi_3, \\
b_2 &=& - \sqrt{3} \gamma_3 ,     & &     d_2 &=&  - \xi_3^\prime .
\end{array}
\eeq

\section{$ K \rightarrow \pi \pi \pi $ Amplitudes.}
\subsection{Leading order}
Employing the lowest order chiral Lagrangian given by \eq{sham}
and \eq{chi-lag}, we can calculate the $ K \rightarrow \pi \pi
\pi$ amplitudes.  The number of the contributing diagrams turn out
to be huge, more than three hundred diagrams. The complete
relevant Feynman rules are too lengthy to be mentioned  but it can
be found in Ref.~\refcite{fey}. Necessary  integral formulas for
calculating Feynman graphs are collected in appendix A along with
many other useful formulas.

The contributing diagrams fall into two classes :
\begin{itemize}
\item{(i)}
Tree diagrams as given in Fig.~(\ref{tree})  turn out to be easily
calculated and their results are polynomial functions of the
external momenta.
\begin{figure}[htbp]
\epsfxsize=8cm
\centerline{\epsfbox{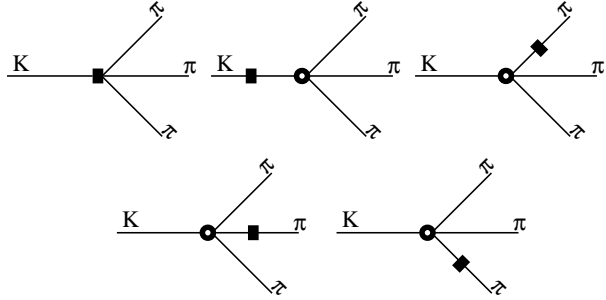}}
\caption{Tree
diagrams for $K\rightarrow \pi\pi\pi$. The black box and circle
indicate the insertion of the weak $\Delta S =1$  and strong
chiral Hamiltonian respectively.}    
\label{tree}
\end{figure}
\item{(ii)} Loop diagrams, The ones in Fig.~(\ref{loop1}) are
tadpole-kind corrections and their results are polynomial
functions of the external momenta. The rest of the  diagrams in
Fig.~(\ref{loop2}) with at most two insertion of strong chiral
Hamiltonian are the most complicated ones.
 Their results are  non polynomial functions of $X$ and $Y$ variables
defined in \eq{defxy}. Hence, we expand the results in Taylor series around
the point $( X = 0, Y= 0)$ which is the center of the Diltaz plot variables.
\begin{figure}[htbp]
\epsfxsize=15cm \centerline{\epsfbox{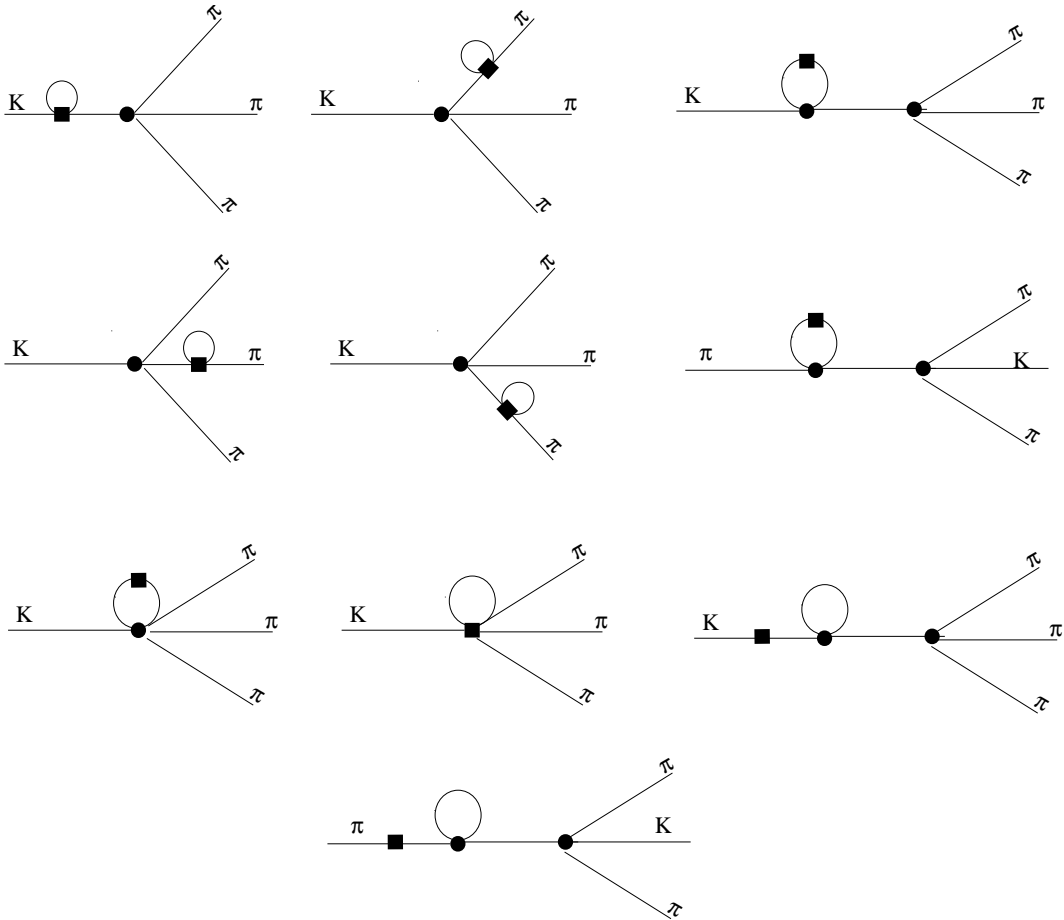}} \caption{
One-loop diagrams of tadpole kind correction for $K\rightarrow \pi
\pi\pi$. The black box and circle indicate the insertion of  the
weak $\Delta S =1$  and strong chiral Hamiltonian respectively. }
\label{loop1}
\end{figure}
\begin{figure}[htbp]
\epsfxsize=8cm
\centerline{\epsfbox{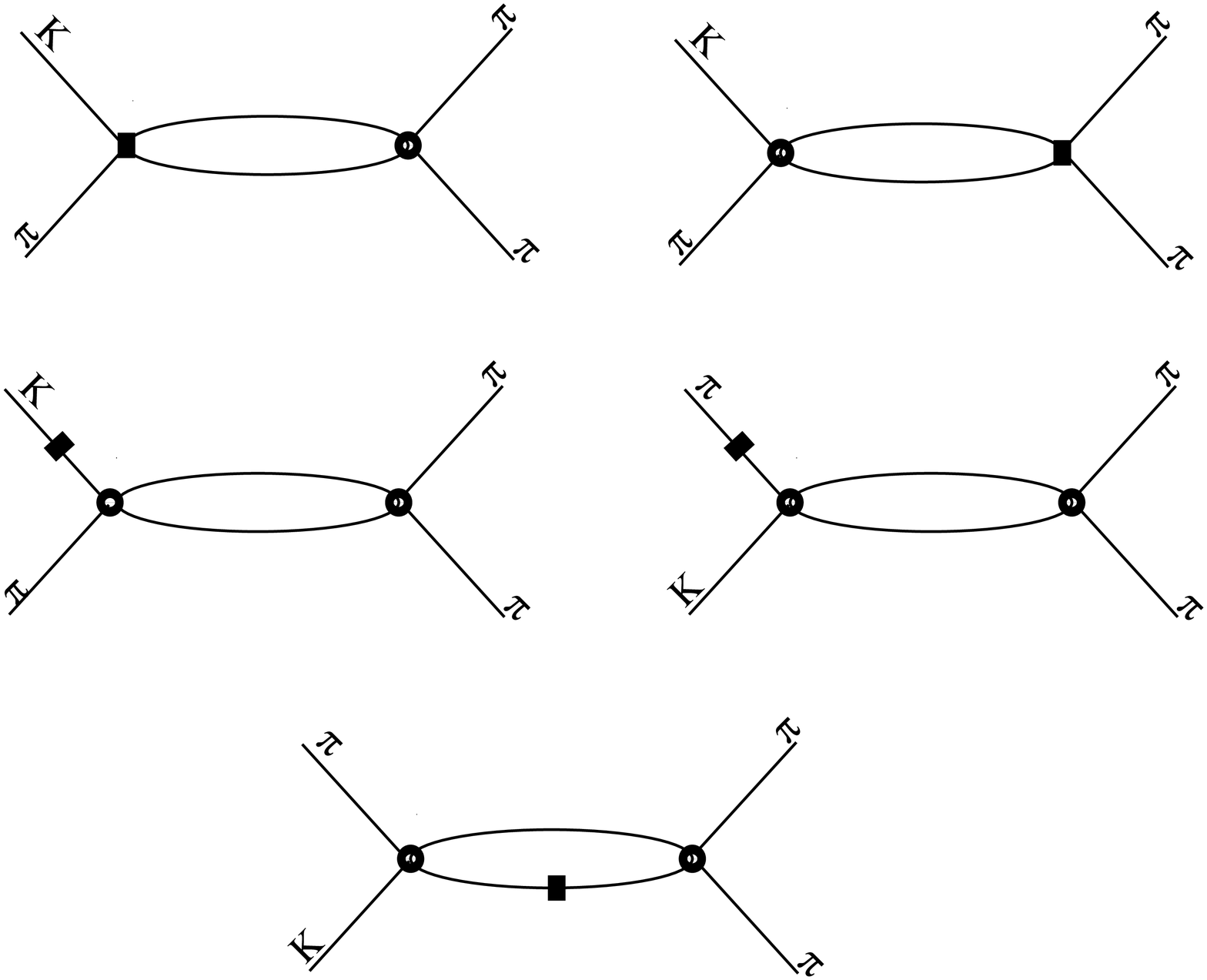}}
\caption{One-loop
diagrams with horizontal loop for $K\rightarrow \pi \pi\pi$. The
black box and circle indicate the insertion of  the weak $\Delta S
=1$ and strong chiral Hamiltonian respectively. }
\label{loop2}
\end{figure}
\end{itemize}

The tree diagrams in Fig.~(1) give the following results.
\bea
A_{++-}(G_a) & = & \frac{2}{3\, f^4} \left( m_K^2 + 3 m_\pi^2 Y \right) \nnu,\\
A_{00+}(G_a) & = & \frac{1}{6\, f^4 }\,   \frac{\left( 2 m_K^4 - 6 m_\pi^4 Y + m_K^2 m_\pi^2
          \left( - 2 + 15 Y\right) \right)}{m_K^2 - m_\pi^2} \nnu, \\
A_{+-0}(G_a) & = & \frac{1}{2\, f^4} \, m_\pi^2 \, \frac{\left( - 3 m_K^2 \left( X + Y\right) +
                     2 m_\pi^2 \left( X + 3 Y \right)\right)}{m_K^2 - m_\pi^2} \nnu, \\
A_{000}(G_a) & = & 0,
\label{gatree}
\eea
\bea
A_{++-}(G_b) & = & \frac{3}{f^4} m_\pi^2 Y ,\nnu \\
A_{00+}(G_b) & = & \frac{3}{2\, f^4}  \frac{ m_K^2 m_\pi^2 Y}{m_K^2 - m_\pi^2} , \nnu \\
A_{+-0}(G_b) & = &  \frac{1}{6\,f^4} \frac{ 2 m_K^4 + 6 m_\pi^4 \left( X + 2 Y \right) -
m_K^2 m_\pi^2 \left( 2 + 9 X + 3 Y\right)} { m_K^2 - m_\pi^2} , \nnu \\
A_{000}(G_b) & = & \frac{m_K^2}{f^4},
\label{gbtree}
\eea
\bea
A_{++-}(G_8) & = & \frac{1}{3\, f^4} \left( - 2 m_K^2 + 3 m_\pi^2 Y \right) ,\nnu \\
A_{00+}(G_8) & = & - \frac{1}{3\, f^4} \left(m_K^2 + 3 m_\pi^2 Y\right) , \nnu \\
A_{+-0}(G_8) & = & \frac{1}{3\, f^4} \left(m_K^2 + 3 m_\pi^2 Y\right) , \nnu \\
A_{000}(G_8) & = & \frac{m_K^2}{f^4}.
\label{g8tree}
\eea

Expressing our leading results in terms of amplitudes with definite isospin selection rule defined in
\eq{rep2}, we find for the $\Delta I = \frac{1}{2}$ amplitudes
\bea
\alpha_1 = \frac{1}{f^4} \frac{m_K^2}{3} \left( g_{\underline{8}} -
 \frac{1}{9}\ g_{\underline{27}}\right),
& & \beta_1 =\frac{1}{f^4} \left(- m_{\pi}^2\right)
\left( g_{\underline{8}} - \frac{1}{9}\ g_{\underline{27}}\right),
\label{leadhalf}
\eea
and the following for the $\Delta I = \frac{3}{2}$
\bea
\alpha_3 = \frac{1}{f^4} \frac{10}{27} m_K^2\ g_{\underline{27}}, & &
\beta_3 = \frac{1}{f^4} \frac{m_{\pi}^2}{m_K^2 - m_{\pi}^2}
\left(5 m_K^2 - 14  m_{\pi}^2 \right) \frac{5}{18}\ g_{\underline{27}}, \nnu \\
\gamma_3 =  \frac{1}{f^4} \frac{m_{\pi}^2}{m_K^2 - m_{\pi}^2}
\left(3 m_K^2 - 2 m_{\pi}^2 \right) \frac{-5}{4 \sqrt{3}} \ g_{\underline{27}}, & &
\label{lead32}
\eea
where the masses $m_{\pi}$, $m_{K}$ and $m_{\eta}$ are taken to be
$0.138$, $0.498$ and $0.548\,$ GeV respectively.

The leading order results in \eqs{leadhalf}{lead32}
agree with those found in  Refs.~\refcite{meson,bij,DI} after taking
care of the used different conventions.

\subsection{Next-to-leading order}
In our present work, only one-loop correction is included, the
other next-to-leading contribution coming from the order $O(p^4)$
counter terms of weak chiral Lagrangian will not be considered
here. In fact, these contributions of weak counter terms can be
computed using the $\chi$QM which is a rather lengthy
calculation. We leave this to a future work.

Other kinds of one-loop corrections are not shown graphically in
Figs.~(\ref{loop1},\ref{loop2}) namely, wave function and f
renormalization. Such kind of corrections originate purely from
the strong sector.

The wave-function renormalization which arise in the chiral perturbation from the direct calculation
of $K \rightarrow K$ and $\pi \rightarrow \pi$ propagators are given at $O(p^2)$ by
\bea
Z_k & = & 1 + i \left[ \frac{1}{4} I_2(m_\pi) + \frac{1}{4} I_2(m_\eta) \right] \ ,\nnu \\
Z_\pi & = & 1 + i \left[ \frac{2}{3} I_2(m_\pi) + \frac{1}{3} I_2(m_k) +
                      \frac{1}{2} I_2(m_K) \right] , \nnu \\
\label{wfc}
\eea
where $I_2(m)$ is
\beq
I_2(m) = \frac{i}{16 \pi^2} m^2 \left[1 - \log({m^2}) + \log({\mu^2})\right]\ .
\label{i2}
\eeq
To include the wave-function renormalization for the amplitude $K \rightarrow \pi \pi \pi $, each tree
level result should be multiplied by the factor $Z_\pi^{\frac{3}{2}} Z_K^{\frac{1}{2}}$.

Concerning the $f$ renormalization; at the lowest order in chiral
perturbation there is no distinction between the decay constants.
After considering one-loop correction, the  physical decay
constants are identified according to the following equation
\bea
f_K = f \left\{1 - \frac{i}{f^2}\left[\frac{3}{8} I_2(m_\pi) +
\frac{3}{4} I_2(m_K) +
\frac{3}{8} I_2(m_\eta)\right]\right\} \ ,\nnu \\
f_{\pi} = f \left\{1 - \frac{i}{f^2}\left[ I_2(m_\pi) +
\frac{1}{2} I_2(m_K)\right]\right\} \ ,\nnu \\
\label{fren}
\eea
where $I_2(m)$ is defined in \eq{i2}. The
inclusion of the $f$ renormalization is made by expressing $1/f^4$
appearing at the tree level in terms of the physical decay
constants $f_{\pi} (=0.0924\;\mbox{GeV})$ and $f_K (=0.1130\;\mbox{GeV})$.
One power of $f$ is eliminated in favor of $f_K$ while the others in
favor of $f_\pi$ through using the relations in \eq{fren}.

In regularizing our loop integrals, we used the dimensional
regularization with a modified minimal subtraction scheme
($\overline{MS}$)\footnote{In this subtraction scheme all the
combination $\frac{1}{\epsilon} - \gamma_E + \log{4\pi}$ is
subtracted, where $\gamma_E$ is the Euler constant.}. For the sake of convenience,
a factor ${10}^{-6}$ is extracted from the results. A part from the tree
level result, the one-loop results after including wave function and $f$
renormalization  are
\bea
A_{++-}(G_a) & = &\frac{1}{f^6} \left[ \left(2835. + 183.3\,i
 +1112.\,\log {{\mu}^2} \right)
                                              +  \right.  \nnu \\
             & &  Y\,\left( 471.1 - 107.4\,i + 217.\,\log {{\mu}^2} \right)
 + \nnu \\
             & &  {X^2}\,\left( 4.259 + 1.317\,i + 1.053\,\log {{\mu}^2} \right) +
\nnu \\
            & &\left.  {Y^2}\,\left( -15.02 - 6.924\,i - 5.455\,\log {{\mu}^2} \right)
\right] , \nnu \\
A_{00+}(G_a) & = &\frac{1}{f^6} \left[ \left(1418. + 91.66\,i +
+555.9\,\log {{\mu}^2}\right) + \right. \nnu \\
             & & Y\, \left( 733.6 + 71.1\,i + 290.6\,\log {{\mu}^2} \right)
+ \nnu \\
             & & {X^2}\,\left(4.352 + 2.937\,i + 0.7177\,\log {{\mu}^2} \right)
+ \nnu \\
             & &\left. {Y^2}\,\left(-14.18 - 10.3\,i - 3.301\,\log {{\mu}^2} \right)
\right] , \nnu \\
A_{+-0}(G_a) & = & \frac{1}{f^6}
\left[\left(648.6+245.5\,\log{{\mu}^2}\right)
                                                            + \right. \nnu \\
             & & X\,\left( -401.5 + 12.1\,i - 169.2\,\log {{\mu}^2} \right)
+ \nnu \\
             & & Y\,\left( -423.7 - 10.79\,i - 163.6\,\log {{\mu}^2} \right)
+ \nnu \\
             & & {X^2}\,\left(-6.657 - 1.486\,i - 2.632\,\log {{\mu}^2} \right)
+ \nnu \\
             & & {Y^2}\,\left( -3.824 - 6.541\,i - 0.7177\,\log {{\mu}^2} \right)
+ \nnu \\
             & & \left. X\,Y\,\left(-18.35 - 9.995\,i - 4.689\,\log {{\mu}^2} \right)
\right] ,\nnu \\
A_{000}(G_a) & = & \frac{1}{f^6} \left[ \left(1946.  + 736.5\,\log
{{\mu}^2} \right) \right. + \nnu\\
             & & {X^2}\,\left(-11.9 - 5.499\,i - 4.306\,\log {{\mu}^2} \right)
+ \nnu \\
             & & \left. {Y^2}\,\left( -35.69 - 16.5\,i - 12.92\,\log {{\mu}^2} \right)
\right] , \nnu \\
\label{galoop}
\eea
\bea
A_{++-}(G_b) & = &\frac{1}{f^6} \left[ \left(1028. +
441.1\,\log {{\mu}^2} \right)
                                                                      + \right.  \nnu \\
             & & Y\,\left( 771.4 - 12.75\,i + 310.5\,\log {{\mu}^2} \right)
+ \nnu \\
             & & {X^2}\,\left(1.764 + 0.492\,i + 0.1914\,\log {{\mu}^2} \right)
+ \nnu \\
            & & \left. {Y^2}\,\left( -29.85 - 12.47\,i - 9.761\,\log {{\mu}^2} \right)
\right] ,\nnu \\
A_{00+}(G_b) & = &\frac{1}{f^6} \left[ \left(513.8 +  220.6\,\log
{{\mu}^2}  \right)
                                                                     +  \right. \nnu \\
             & & Y\, \left( 432.4 - 23.54\,i + 196.7\,\log {{\mu}^2} \right)
                                                                           + \nnu \\
             & & {X^2}\,\left( 1.27 + 1.756\,i - 0.1435\,\log {{\mu}^2} \right)
+ \nnu \\
             & &  \left. {Y^2}\,\left(-16.09 - 10.77\,i - 4.163\,\log {{\mu}^2} \right)
\right] , \nnu \\
A_{+-0}(G_b) & = & \frac{1}{f^6} \left[ \left( 1555. + 91.66\,i
+581.9\,\log {{\mu}^2} \right)
                                                                       + \right. \nnu \\
             & &X\,\left( -401.3 + 12.1\,i - 169.1\,\log {{\mu}^2} \right)
+ \nnu \\
             & &Y\,\left( -123.1 + 83.85\,i - 69.95\,\log {{\mu}^2} \right)
+ \nnu \\
             & & {X^2}\,\left( -3.575 - 0.3046\,i - 1.77\,\log {{\mu}^2} \right)
+ \nnu \\
             & & {Y^2}\,\left( -1.916 - 6.072\,i + 0.1435\,\log {{\mu}^2} \right)
+ \nnu \\
             & & \left. X\,Y\,\left( -18.35 - 9.995\,i - 4.689\,\log {{\mu}^2}  \right)
\right],  \nnu \\
A_{000}(G_b) & = & \frac{1}{f^6} \left[\left(4665. + 275.\,i +
1746.\,\log {{\mu}^2} \right)
                                                                      \right. +  \nnu\\
             & & {X^2}\,\left( -6.321 - 3.493\,i - 2.584\,\log {{\mu}^2} \right)
+ \nnu \\
             & &\left. {Y^2}\,\left(-18.96 - 10.48\,i - 7.751\,\log {{\mu}^2} \right)
\right] ,
\label{gbloop}
\eea
\bea
A_{++-}(G_8) & = &\frac{1}{f^6} \left[ \left(-1808. -
183.3\,i
 - 670.6\,\log {{\mu}^2} \right) +  \right.  \nnu \\
             & &Y\,\left( 300.3 + 94.64\,i + 93.49\,\log {{\mu}^2} \right)
+ \nnu \\
             & &{X^2}\,\left( -2.495 - 0.8252\,i - 0.8612\,\log {{\mu}^2} \right)
+ \nnu \\
            & & \left. {Y^2}\,\left(-14.82 - 5.551\,i - 4.306\,\log {{\mu}^2} \right)
\right] , \nnu \\
A_{00+}(G_8) & = &\frac{1}{f^6} \left[ \left(-903.9 - 91.66\,i -
   335.3\,\log {{\mu}^2} \right)
                                                                           +\right. \nnu \\
             & &  Y\,\left( -301.2 - 94.64\,i - 93.85\,\log {{\mu}^2} \right)
+ \nnu \\
             & & {X^2}\,\left( -3.082 - 1.181\,i - 0.8612\,\log {{\mu}^2} \right)
+ \nnu \\
             & &  \left. {Y^2}\,\left(-1.908 - 0.4689\,i - 0.8612\,\log {{\mu}^2} \right)
\right] , \nnu \\
A_{+-0}(G_8) & = & \frac{1}{f^6} \left[ \left(906.6 + 91.66\,i +
336.4\,\log {{\mu}^2} \right)
                                                                              + \right. \nnu \\
             & &  X\, \left( 0.2895 + 0.1209\,\log {{\mu}^2} \right)
             +  \nnu \\
             & & Y\,\left( 300.6 + 94.64\,i + 93.61\,\log {{\mu}^2} \right)
+ \nnu \\
             & & {X^2}\,\left(3.082 + 1.181\,i + 0.8612\,\log {{\mu}^2} \right)
+ \nnu \\
             & &   \left. {Y^2}\,\left(1.908 + 0.4689\,i + 0.8612\,\log {{\mu}^2}\right)
\right] ,\nnu \\
A_{000}(G_8) & = & \frac{1}{f^6} \left[ \left(2720. + 275.\,i +
1009.\,\log {{\mu}^2} \right)
                                                                                + \right.  \nnu\\
             & & {X^2}\,\left( 5.577 + 2.007\,i + 1.722\,\log {{\mu}^2}\right)
+ \nnu \\
             & & \left. {Y^2}\,\left(16.73 + 6.02\,i + 5.167\,\log {{\mu}^2} \right)
\right] ,
\label{g8loop}
\eea
where $\mu$ is the renomalization
scale measured in units of GeV and $i = \sqrt{-1}$.

As a check of the calculations, one can find different ways for
determining the parameters $ a_c , b_c,\cdots$ etc. defined
in \eq{rep1}  which should give the same result. As an example
$a_c$ can be determined from both $A_{++-}$ and $A_{00+}$ leading
to the same result. Similar patterns hold for the rest of the
parameters.

Another check comes from observing that the term \beq \Tr \left(
\lambda^3_2 \Sigma^{\dag} D_\mu\Sigma \right) \Tr \left(
\lambda^1_1 \Sigma^{\dag} D^\mu \Sigma \right) - \Tr \left(
\lambda^3_1 \Sigma^{\dag} D_\mu \Sigma \right) \Tr \left(
\lambda^1_2 \Sigma^{\dag} D^\mu  \Sigma \right) \ , \eeq transform
as an octet $(8_L \times 1_R)$.  Computing Feynman diagrams for
each part separately and then subtracting them, the result should
be the same as of the octet part of the weak chiral Lagrangian. In
fact, these two tests  constitute  non trivial checks for the
performed calculations and our calculations did pass them successfully.
There is an additional check which holds for the
imaginary parts of the amplitudes and is explained in appendix~A.

Expressing our one-loop results in terms of dimensionless amplitudes with
definite isospin selection rules (apart from the tree level
results  which are given by \eqs{leadhalf}{lead32},
we get for the octet
\bea
\alpha_1^{(8)} &=& 12.402 + 1.256\,i + 4.601\,\log ({{\mu}^2})\ , \nnu \\
\alpha_3^{(8)} &=& 0.025 +0.010\,\log ({{\mu}^2})\ ,  \nnu \\
\beta_1^{(8)} & = &-4.122 - 1.297\,i -1.284\,\log ({{\mu}^2})\ , \nnu \\
\beta_3^{(8)} & = &0.001 + 0.00055\,\log ({{\mu}^2})\ ,  \nnu \\
\zeta_1^{(8)} & = & 0.076 + 0.028\,i +0.024\,\log ({{\mu}^2}) \ , \nnu \\
\xi_1^{(8)} &=& -0.050 - i\,0.021 -0.012\, \log{(\mu^2)} \ , \nnu \\
\gamma_3^{(8)} & = & 0.003 + 0.0014\,\log({{\mu}^2})\ ,
\label{nleadhalf}
 \eea
 and for the 27-plet
 \bea
\alpha_1^{(27)}&=&-4.250 - 0.134\,i -1.995\,\log ({{\mu}^2}) \ , \nnu \\
\alpha_3^{(27)} &=& 31.494 + 1.396\,i +12.216\,\log ({{\mu}^2})\ , \nnu \\
\beta_1^{(27)} & =& 1.104 + 0.144\,i +0.522\,\log ({{\mu}^2})\ , \nnu \\
\beta_3^{(27)} & =&  4.455 - 1.195\,i +1.932\,\log ({{\mu}^2})\ , \nnu \\
\gamma_3^{(27)} &=& -7.941 + 0.239\,i -3.346\,\log ({{\mu}^2}) \ , \nnu \\
\zeta_1^{(27)} & =&  -0.006 - i\, 0.003 -
   0.0004\,\log{(\mu^2)} \ , \nnu \\
\zeta_3^{(27)} & =& 0.095 +i\, 0.048 + 0.037\,\log{(\mu^2)} \ , \nnu \\
\xi_1^{(27)} &=&  0.009 + i\,0.002 +
   0.003\,\log{(\mu^2)} \ , \nnu \\
\xi_3^{(27)} &=&  -0.063 + i\, 0.024 -
   0.033\,\log{(\mu^2)} \ , \nnu \\
\xi_3^{\prime (27)} &=&  0.314 + i \, 0.171 + 0.080\, \log{(\mu^2)}. \nnu \\
\label{nleading32}
\eea
The results in \eqs{nleadhalf}{nleading32} are made to be
dimensionless through multiplying by the factor ${f_\pi^2}$.
Then the full one-loop contributions to the isospin amplitudes are obtained by
setting $A_{(i)}~=~(\frac{g_{\underline{8}}}{f_\pi^2})
A_{(i)}^{(8)} +
 (\frac{g_{\underline{27}}}{f_\pi^2}) A_{(i)}^{(27)} $.

The values of the isospin amplitudes  depend on the scheme and
scale of renormalization. For a meaningful comparison we restrict
ourselves to the imaginary part which are scale and
subtraction-scheme independent, leaving the comparison of the real parts
for the final fit to the experimental results.
\begin{table}[hbt]
\tbl{Comparison of the imaginary parts of the isospin amplitudes.}
{\begin{tabular}{@{}ccccccccc@{}}
\hline
\\
& $\alpha_1^{(8)}$ &$\beta_1^{(8)}$&$\zeta_1^{(8)}$ &$\xi_1^{(8)}$&$\alpha_1^{(27)}$&
 $\beta_1^{(27)}$&$\zeta_1^{(27)}$&$\xi_1^{(27)}$\\
\\
\hline
Kambor \etal &1.00& 0.50& 0.019& 0.00&- 0.67&0.33 &- 0.012&  0 \\
Our calculations &1.26&-1.30& 0.028 & - 0.021 & -0.42 & 0.432 &-0.009& 0.006 \\
\hline
\\
& $ \alpha_3^{(27)}$ &$\beta_3^{(27)}$&$\gamma_3^{(27)}$ &$\zeta_3^{(27)}$ &$\xi_3^{(27)}$
& $\xi_3^{\prime (27)}$& & \\
\\
\hline
Kambor \etal & 6.70 & -2.19& 1.12 &  0.066 & 0 & 0.43 & & \\
Our calculation & 4.2 & - 3.585 & 0.717 & 0.144 & 0.072 & 0.513 & & \\
\hline
\end{tabular}
\label{complex}
}
\end{table}

The results in  Table~(\ref{complex}) indicate that there is a
reasonable agreement of our calculations and those of Ref.~\refcite{meson}
for most of the imaginary parts of the isospin amplitudes.
The rest of the parameters $\beta_1^{(8)}, \beta_3^{(27)}$ and
$\gamma_3^{(27)}$ disagree.

In Ref.~\refcite{wise}, the imaginary part of the amplitude of the channel
$K^+\rightarrow \pi^+ \pi^+ \pi^-$ was calculated  for the octet part
of the weak chiral Lagrangian leading to the result
\bea
\Im\left[A\left(K^+\rightarrow \pi^+ \pi^+ \pi^-\right)\right]= \hspace{9cm} \nnu \\
  \frac{-1}{192\ \pi\ f_\pi^6}\ \left[  6\ \sqrt{1 - \frac{4 m_\pi^2}{s_3}} \left(m_K^2 + m_\pi^2
- s_1 - s_2 \right) \left(2 m_\pi^2 - s_1 - s_2\right) \right.   \nnu \\
   +  2\ \sqrt{1 - \frac{4 m_\pi^2}{s_1}} \left(9 m_\pi^4 + m_K^2 \left(2 m_\pi^2 + s_1\right)
 + s_1\left(5 s_1 + s_2\right) - m_\pi^2\left(11 s_1 + 4 s_2\right)\right) \nnu \\
  + \left.  2\ \sqrt{1 - \frac{4 m_\pi^2}{s_2}} \left(9 m_\pi^4 + m_K^2 \left(2 m_\pi^2 + s_2\right)
 + s_2\ \left(5 s_2 + s_1\right) - m_\pi^2\left(11 s_2 + 4 s_1\right)\right)
\right] \nnu \\
\label{wiseim}
\eea
Expanding the result in \eq{wiseim} in  terms
of $X$ and $Y$ up to quadratic terms, we get
\beq
\Im\left[f_\pi^2\, A(K^+\rightarrow \pi^+ \pi^+ \pi^-)\right] = -2.51  + 1.2973\, Y -
0.0113109\, X^2 - 0.0760878\, Y^2. \label{nwise}
\eeq
The imaginary parts of the dimensionless  isospin amplitudes determined from
\eq{nwise} are \beq \alpha_1^{(8)} = 1.26,\ \beta_1^{(8)} =
-1.30,\ \zeta_1^{(8)}=0.028,\ \xi_1^{(8)}=-0.021, \eeq which are
in a full agreement with our result in Table~(\ref{complex}). A
similar agreement holds for the twenty-seven part after paying
some caution to the convention followed in Ref.~\refcite{wise}. Our
analytical result of the $\Im\left[A\left(K^+\rightarrow \pi^+
\pi^+ \pi^-\right)\right]$ , for both octet and twenty seven parts
of the Lagrangian, with those of Ref.~\refcite{wise} can be found in
appendix~A.

It is difficult to point out the source of discrepancies between
the two results mentioned above, in particular, since the details of
work Ref.~\refcite{meson} were lost as stated before in the introduction.
However, as shown in the appendix, some theoretical identities relating the
imaginary parts of the amplitudes are satisfied for our results
but fail in Ref.~\refcite{meson}. Moreover, our results agree with
other previously calculated ones in Ref.~\refcite{wise}. Thus,
we suspect that the discrepancies might be attributed to some
missing diagrams in the calculations of Ref.~\refcite{meson}.

\section{ Input parameters}
Before discussing the fit of the isospin amplitudes
with the experimental data, we should consider the relevant input
parameters for the chiral quark model.

The value of the input parameters $M$, the gluon condensate
$\langle \frac{\alpha_s}{\pi} G G \rangle$ and the quark condensate  $\langle \overline{q} q \rangle$
 are chosen such that to fit the $\Delta I = 1/2 $ selection rule characterizing the
$I = 0$ and $I = 2$ amplitude of the non-leptonic Kaon decay ($K\rightarrow 2 \ \pi$).

Such a fit was done in Ref.~\refcite{V} for the complete $O(p^4)$ calculation in the framework of
$\chi$QM.
Since, in our present work, we stop at $O(p^2)$ order while including the meson loop
correction, we redid the fit in order to determine the appropriate values of
the parameters. The resulting values of the input parameters which fit the
$\Delta I =1/2$ rule up to $20 \%$ are
\bea
\langle \frac{\alpha_s}{\pi} G G \rangle  & = & (358 ^{+7}_{-12}\,  \mbox{MeV})^4 \ , \nnu \\
\langle \overline{q} q \rangle & = & - (260 ^{+43}_{-100}\, \mbox{MeV})^3 \ , \nnu \\
M & = & 200 \pm 20\,  \mbox{MeV}.
\label{input}
\eea

Let us focus on the coupling constants $g_{\underline{8}}$ and
$g_{\underline{27}}$ defined in \eq{ourcoupling}.  For convenience we divide them by
$f_{\pi}^2$ to produce dimensionless quantities measured in units of $10^{-8}$.
For our central values of the parameters $(M,\langle \frac{\alpha_s}{\pi} G G \rangle,
\langle \overline{q} q \rangle )$, the couplings take the following values
\beq
\frac{g_{\underline{8}}}{f_{\pi}^2} = 3.42 \ , \
\frac{g_{\underline{27}}}{f_{\pi}^2} = -0.382.
\eeq
These values show clearly the dominance of the octet coupling $(g_8)$, which
is necessary for producing the $\Delta\, I = {1\over 2}$ selection rule.

\begin{figure}[hbtp]
\epsfxsize=10cm \centerline{\epsfbox{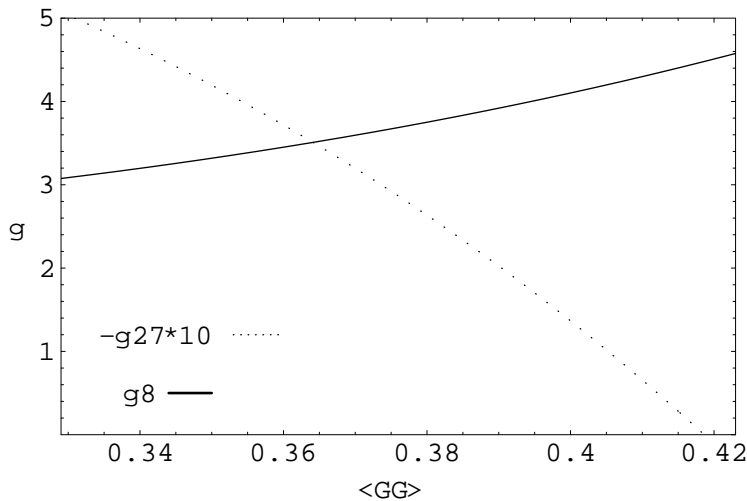}}
\caption{The solid (dotted) line
represents the coupling $g_{\underline{8}}/f_{\pi}^2$ ($|g_{\underline{27}}/f_{\pi}^2|\times 10$)
as a function of the gluon condensate. $\langle \overline{q} q \rangle$ and M
are fixed at their central values in \eq{input}.}
\label{g827}
\end{figure}

We illustrate the dependence of the couplings
on the input parameters in Fig.~(\ref{g827}). For convenience we kept  $\langle \overline{q} q \rangle$ and M
fixed at their central values in \eq{input}, while we varied the gluon condensate in the following range
\beq
\langle \frac{\alpha_s}{\pi} G G
\rangle = (329-423\ \mbox{MeV})^4,
\eeq
As evident from Fig.~(\ref{g827}), increasing the gluon condensate tends to enhance the coupling
$g_{\underline{8}}$ while suppressing  the coupling $g_{\underline{27}}$.
It is only the coupling $g_{\underline{8}}$ which depends on the quark condensate and $M$.
It turns out that the coupling $g_{\underline{8}}$ is enhanced by increasing the quark condensate
or by reducing $M$. In the foregoing discussion, the renormalization scale $\mu$ was fixed at $0.8\ \mbox{GeV}$,
and the scheme is taken to be the HV one. The $\g$-scheme and scale dependence of our final results are discussed
at the end of the next section.

It is worth  mentioning that the phenomenological determination of the quark and gluon condensate
are a complicated issue and the literature offers different determinations of them. To fix the ideas, the
condensate entering our computation are identified with those obtained by fitting the experimental data
using the  QCD sum rules (QCD-SR) methods  or the lattice computations.

Regarding the gluon condensate, it was first  determined by
Shifman, Vainshtein, and Zakkharov (SVZ) using
QCD-SR.~\cite{srr1}\cdash\cite{srr3} Later, lattice QCD was used for this
determination.~\cite{lat1}\cdash\cite{lat3}
We prefer to take the value of the gluon condensate in the range
\beq
\langle \frac{\alpha_s}{\pi} G G
\rangle = (376\pm 10\ \mbox{MeV})^4,
\label{rg1}
\eeq
which encompasses the result of QCD-SR
analysis.~\cite{sr2}\cdash\cite{sr4}
The corresponding ranges provided by the lattice
calculations in Refs.~\refcite{latt2,latt3}
are respectively,
\beq
\langle \frac{\alpha_s}{\pi} G G
\rangle = (460\pm 21\ \mbox{MeV})^4,\, \mbox{and}\,\,
\langle \frac{\alpha_s}{\pi} G G
\rangle = (651\pm 40\ \mbox{MeV})^4,\\
\label{rg2}
\eeq
and they  are larger than the estimates provided by QCD-SR.

For the quark condensate, we take the range
\beq
-(200\, \mbox{MeV})^3 \le \langle \bar{q}\,q\rangle \le -(280\,\mbox{MeV})^3
\label{qqr}
\eeq
in order to include the central values and errors of the
QCD-SR\cite{es1} and the lattice estimate\cite{ess2}\cdash\cite{ess3}.

As to the constituent quark mass $M$, we consider the range
\beq
M\approx 200 - 250 \mbox{MeV},
\label{rm}
\eeq
which is consistent with the estimate based on processes involving
mesons.~\cite{bij5} Such a value is smaller than the value
$M\approx 330\,\mbox{MeV}$ often quoted from baryon physics.

\section{Comparison with the experimental data}
In comparing the $ K \rightarrow \pi \pi \pi $ amplitudes with the
experimental data, we use the expansion given in \eq{rep2}.
The analysis of the parameters $\alpha_1 , \alpha_2, \cdots etc.$
was first made by Devlin and Dickey  in the late
seventies.~\cite{dev} In the early nineties  the analysis has been redone by
Kambor \etal.\cite{meson} using recent data and including the
$\Delta I = 3/2$ quadratic slope parameters $\zeta_3$ and $\xi_3^\prime$
in \eq{rep2}. More recently \cite{bij}  a full fit has been done
using recent data in Ref.~\refcite{pdg}. The fitted values for the
parameters $ \alpha_1 , \alpha_2, \cdots etc.$ are displayed in
Table~(\ref{isoexpamp}).
\begin{table}[h]
\tbl{Determination of $ K \rightarrow 3 \pi $ isospin
amplitudes, in units of $10^{-8}$.}
{\begin{tabular}{@{}|c|c|c|c|@{}}
\hline
Quantity & Ref.\cite{dev} & Ref.\cite{meson} & Ref.\cite{bij}\\
\hline $\alpha_1$ & $91.4\pm0.24$ & $91.71\pm0.32$ &$93.16\pm
 0.36$
\\
\hline
$\alpha_3$ & $-7.14\pm 0.36$ & $-7.36\pm0.47$ &
$-6.72\pm0.46$ \\
\hline
$\beta_1$ & $-25.83\pm0.41$ & $-25.68\pm 0.27$ & $-27.06\pm 0.43$ \\
\hline
$\beta_3$ & $-2.48\pm0.48$ & $-2.43\pm 0.41$ & $-2.22\pm 0.47$ \\
\hline
$\gamma_3$ & $2.51\pm 0.36$ & $2.26\pm 0.23$ & $2.95\pm 0.32$ \\
\hline
$\zeta_1$ & $-0.37\pm 0.11$ & $-0.47\pm 0.15$ & $-0.40\pm 0.19 $ \\
\hline
$\zeta_3$ &$ \--$ & $-0.21 \pm 0.08 $ & $-0.09 \pm 0.10 $ \\
\hline
$\xi_1$ & $-1.25\pm 0.12 $ & $-1.51\pm 0.30 $ & $-1.83\pm 0.30$ \\
\hline
$\xi_3$ & $\--$ & $-0.12\pm 0.17$ & $-0.17\pm 0.16$ \\
\hline
$\xi_3^\prime$ & $\--$ & $-0.21\pm 0.51$ & $-0.56\pm 0.42$ \\
\hline
\end{tabular}
\label{isoexpamp}
}
\end{table}

As discussed in Refs.\refcite{I}--\refcite{VI}, it was shown that within
the frame work of $\chi$QM , the $\Delta I = 1/2$ rule in the case
of $K\rightarrow 2 \pi$ decay can be reproduced. Here we show how
this can be achieved for the case $K\rightarrow3 \pi$. Since
our results depend on the following parameters Gluon $\langle
\frac{\alpha_s}{\pi} G G \rangle$ , quark $\langle \overline{q} q
\rangle$ condensate and the constituent mass M, we choose the
values which fit the $\Delta I = 1/2$ rule for the $K\rightarrow 2
\pi$ determined by \eq{input} as mentioned in section~7. We show
the numerical results for the isospin amplitudes at the central
value of this range in Table~(\ref{cfit}).
\begin{table}[h]
\tbl{The isospin amplitudes, in units of $10^{-8}$ at the
central value of the range given by \eq{input}.}
{\begin{tabular}{@{}|c|c|c|c|c|c|c|c|c|c|@{}}
\hline $\alpha_1$ & $\alpha_3$&$\beta_1$ & $\beta_3$ & $\gamma_3$
&$\zeta_1$ & $\zeta_3$ & $\xi_1$ & $\xi_3$ & $\xi_3^\prime$ \\
\hline 64.15 & -13.23 & -18.8 & -2.18 & 4.02 & 0.23 & -0.03 &
-0.16 & 0.018 & -0.11 \\
\hline
\end{tabular}
\label{cfit}
}
\end{table}
The results are clearly off by a factor of two for $\alpha_3$ while
being in a reasonable agreement for the rest of the linear slope parameters.

In order to improve the result, there should be a suppression of
$ \alpha_3$ and $\gamma_3$ which can be achieved by lowering the
coupling $g_{\underline{27}}$. This  can be done by increasing the
gluon condensate. We find the best fit corresponds to the
following values;
\bea
\langle \frac{\alpha_s}{\pi} G G \rangle  & = & (376\  \mbox{MeV})^4 \ , \nnu \\
\langle \overline{q} q \rangle & = & - (260\  \mbox{MeV})^3 \ , \nnu \\
M & = & 200\   \mbox{MeV}.
\label{bestinput}
\eea
which for the gluon is slightly out of the range specified by \eq{input}, but it
is still safe for fit of the $\Delta I =1/2$ in the $K\rightarrow
\pi\pi$ decay within $30\%$.

Here in Table~(\ref{bfit})  we list our results together with the
ones obtained in Ref.~\refcite{bij} corresponding to the case of
neglecting higher order counter terms for the sake of a meaningful
comparison.
\begin{table}[h]
\tbl{The value of $ K \rightarrow 3 \pi $ isospin amplitudes,
in units of $10^{-8}$ for case of neglecting higher order counter
term.}
{\begin{tabular}{@{}|c|c|c|@{}}
\hline
Quantity & Ref.\cite{bij} & Our results\\
\hline $\alpha_1$ & $59.4$ &$68.55$
\\
\hline $\alpha_3$ & $-6.5$ &$-9.91$ \\
\hline
$\beta_1$ & $-21.9$ & $-20.11$ \\
\hline
$\beta_3$ & $-1.0$ & $-1.64$ \\
\hline
$\gamma_3$ & $2.5$ & $3.02$ \\
\hline
$\zeta_1$ & $0.26$ & $0.24 $ \\
\hline
$\zeta_3$ & $-0.01$ & $-0.02 $ \\
\hline
$\xi_1$ & $-0.46$ & $-0.17$ \\
\hline
$\xi_3$ & $-0.01$ & $0.01$ \\
\hline
$\xi_3^\prime$ & $-0.06$ & $-0.08$ \\
\hline
\end{tabular}
\label{bfit}
}
\end{table}
As it can be seen there is a reasonable agreement within $30\%$ for
the parameters $\alpha_1, \beta_1 ,\gamma_3, \zeta_1 $ and
$\xi_3^\prime$. Even if we compare with the full fit \cite{bij} listed in
Table~(\ref{isoexpamp}), there is a reasonable agreement for the
linear slope parameters $\alpha_1, \beta_1 , \beta_3$, and
$\gamma_3$ except for $\alpha_3$ which agree within $50\%$. As to
quadratic slope parameters only $\zeta_3$ is compatible with the
fit while the others still need further improvement. The result are
expected to be improved after including the higher order counter
terms. These kind of correction can be calculated in the chiral
quark model, which are rather lengthy and will be the subject of
future work. Another feature which is worthy to be discussed namely
the $\g$-scheme and scale dependence of the amplitudes. As a direct
measure of the scale dependence we define
\beq
\Delta_\mu A \equiv 2 \:\left| \frac{A (0.8 \: \mbox{GeV}) -
 A (1.0 \: \mbox{GeV})}
 {A ( 0.8 \: \mbox{GeV}) + A(1.0 \: \mbox{GeV}) }\right| \, ,
 \eeq
while the difference between the HV and NDR results is quantified
by
\beq
\Delta_{\gamma_5} A_i
 \equiv 2 \:\left| \frac{A_i^{\rm NDR} - A_i^{\rm HV}}{A_i^{\rm NDR} +
 A_i^{\rm HV}}\right| \, .
 \eeq
As evident from Table~(\ref{sscheme}) the scale dependence of the
isopin amplitudes is within $20\%$ excluding $\zeta_3$ and $\xi_3$
for which the scale dependence turns out to be about $30\%$. As to the
$\g$-scheme dependence is not large for the amplitudes and remains
below $10\%$. The overall dependence is not large and still under
control.
\begin{table}[h]
\tbl{The scale and $\g$ scheme dependence of the isospin amplitudes
$A$ (in units of $10^{-8}$). The amplitudes are computed for the values of the
parameters given by \eq{bestinput}. }
{\begin{tabular}{@{}|c|c|c|c|c|c|c|c|c|c|c|c|@{}}
\hline
$\mbox{A}$&\m{3}{|c|}{$\mu=0.8\, \mbox{GeV}$}&\m{3}{|c|}{$\mu=0.9\,
\mbox{GeV}$}&\m{3}{|c|}{$\mu=1.0\, \mbox{GeV}$}& \m{2}{|c|}{$\De_\mu A$}\\
\cline{2-12}
&$\mbox{HV}$&$\De_{\g}A$ &$\mbox{NDR}$&$\mbox{HV}$&$\De_{\g}A$ &$\mbox{NDR}$
&$\mbox{HV}$&$\De_{\g}A$&$\mbox{NDR}$&$\mbox{HV}$ &$\mbox{NDR}$ \\
\cline{2-12}
$\alpha_1$ &$68.55$&$4\%$&$66.05$ &$61.72$&$3\%$ &$59.66$ &$57.77
$&$3\%$ &$56.16$ &$17\%$ &$16\%$\\
\hline
$\alpha_3$ &$-9.91$&$7\%$ &$-10.70$ &$-11.18$&$6\%$ &$-11.91$ &$-12.30
$&$6\%$ &$-13.0$ &$22\%$ &$19\%$ \\
\hline
$\beta_1$ &$-20.11$&$4\%$ &$-19.37$ &$-18.05$&$3\%$ &$-17.44$ &$-16.84
$&$3\%$ &$-16.37$ &$18\%$ &$17\%$\\
\hline
$\beta_3$ &$-1.64$&$8\%$ &$-1.77$ &$-1.85$&$6\%$ &$-1.97$ &$-2.03
$&$5\%$ &$-2.14$ &$21\%$ &$19\%$\\
\hline
$\gamma_3$ &$3.02$&$8\%$ &$3.26$ &$3.38$&$6\%$ &$3.60$ &$3.69
$&$6\%$ &$3.9$ &$20\%$ &$18\%$\\
\hline
$\zeta_1$ &$0.245$&$4\%$ &$0.235$ &$0.224$&$3\%$ &$0.217$ &$0.213
$&$3\%$ &$0.207$ &$14\%$ &$13\%$\\
\hline
$\zeta_3$ &$-0.022$&$9\%$ &$-0.024$ &$-0.026$&$7\%$ &$-0.028$ &$-0.030
$&$3\%$ &$-0.031$ &$31\%$ &$25\%$\\
\hline
$\xi_1$ &$-0.168$&$4\%$ &$-0.162$ &$-0.151$&$3\%$ &$-0.146$ &$-0.142
$&$3\%$ &$-0.138$ & $17\%$ &$16\%$\\
\hline
$\xi_3$ &$0.014$&$7\%$ &$0.015$ &$0.017$&$6\%$ &$0.018$ &$0.019
$&$5\%$ &$0.020$ &$30\%$ &$29\%$\\
\hline
$\xi_3^\prime$ &$-0.080$&$7\%$ &$-0.086$ &$-0.089$&$5\%$ &$-0.094$ &$-0.096
$&$6\%$ &$-0.102$ &$18\%$ &$17\%$\\
\hline
\end{tabular}
\label{sscheme}
}
\end{table}

Finally one can say, with caution, that chiral quark model
provides a coherent right picture for the K-meson physics not only
for $K\rightarrow \pi \pi$ decay as shown in
Refs.~\refcite{I}--\refcite{VI}
but also for case $K\rightarrow \pi \pi \pi$.

\section{Conclusion}
In this paper we have calculated the $K\rightarrow \pi\pi\pi$
amplitudes by including the one loop correction in chiral
perturbation theory in the framework of chiral quark model. Our
calculations are independent of what have been carried out in
Refs.~\refcite{meson,bij}. The results for the chiral loop corrections are
presented in a way to be more manageable and easier for use.
Further more we derive a useful new identity for the imaginary part of
the $K\rightarrow \pi\pi\pi$ amplitudes, the details are left for
the appendix.

Our results for the isospin amplitudes are compared with the
recent fit done by,~\cite{bij} we find a reasonable agreement for
the linear slope parameters except $\alpha_3$. As to the quadratic
slope parameters only $\zeta_3$ is compatible with the recent fit.
However restricting the comparison to the fit while neglecting
higher order counter terms leads to a reasonable agreement for
almost all the slope parameters.

The result can be further improved by including higher order
counter terms in the strong and weak chiral lagrangian, which can
be calculated in the frame work of chiral quark model. However the
calculations are rather lengthy and will be the subject of future
work.

\section*{Acknowledgements}
I would like to thank N. Paver, S. Bertolini and M. Fabbrichesi for many
useful discussions. Also I would like to acknowledge the hospitality of ICTP and INFN where the
major part of this work has been done.

\appendix

\section{Integral Formulas, Conventions and Imaginary Part of the Amplitudes }

In this appendix A we collect all the integral formulas necessary for our calculations and summarize
the conventions used in writing the weak chiral Lagrangian.
Moreover we present some of our calculated analytical results for
the imaginary part of the amplitudes and prove some identities which should be
satisfied by the results.

All the loop integrals are evaluated by using dimensional regularization assuming the space-
time dimension  $d = 4 - 2\,\epsilon $. In subtracting the divergences we follow
the modified minimal subtraction scheme ($\overline{MS}$), which means that not only the pole term
$\frac{1}{\epsilon}$ is subtracted but the whole combination
$(\frac{1}{\epsilon} - \gamma_E + \log{4 \pi})$.

The necessary formulas  for loop integrals are:
\beq
\begin{array}{rll}
f_1\left(x,m_1,m_2\right) & = & m_2^2 - x \left(m_2^2 - m_1^2\right), \nnu \\
f_2\left(x,p^2,m_1,m_2\right) & =&
m_2^2 - x\left(1-x\right) p^2 - x \left(m_2^2 - m_1^2\right),
\nnu \\
f_3\left(x,y,p^2,m_1,m_2,m_3\right) & =& m_2^2 -x\left(1-x\right) p^2 -
x \left(m_2^2 - m_1^2\right) -y\left(m_2^2 - m_1^2\right). \nnu \\
\end{array}
\eeq
The finite parts for the required loop integrals are:
{\footnotesize
\bea
&\int  \frac{d^d q}{\left(2 \pi\right)^d} \, \frac{1}{q^2 -
m^2}=\hspace{9cm}\nnu \\
&\frac{i}{16 \pi^2} m^2 \left(1 - \ln{m^2} + \ln{\mu^2}\right),\nnu
\eea
\vspace{-5mm}
\bea
&\int  \frac{d^d q}{\left(2 \pi\right)^d} \, \frac{1}{\left[(q + p)^2 -
 m_1^2 \right] (q^2 - m_2^2)} =\hspace{9cm}\nnu\\
&\frac{i}{16 \pi^2}\left[-\int_0^1 dx \ln{f_2\left(x,p^2,m_1,m_2\right)} +
\ln{\mu^2}\right], \nnu
\eea
\vspace{-5mm}
\bea
& \int  \frac{d^d q}{\left(2 \pi\right)^d} \,  \frac{ (q \cdot p_1) }
{\left[(q + p)^2 - m_1^2 \right] (q^2 - m_2^2)}=\hspace{9cm}\nnu\\
 & \frac{i}{16 \pi^2}
\left[\int_0^1 dx\,\, x\,\, p\cdot p_1 \ln{f_2\left(x,p^2,m_1,m_2\right)}
- p\cdot p_1 \ln{\mu^2}\right],  \nnu
\eea
\vspace{-5mm}
\bea
&\int \frac{d^d q}{\left(2 \pi\right)^d}  \,  \frac{  (q \cdot p_1)  (q \cdot p_2)}
{\left[(q + p)^2 - m_1^2 \right] (q^2 - m_2^2)}=\hspace{8cm}\nnu\\
&\frac{i}{32 \pi^2}\, p_1\cdot p_2
\int_0^1 dx\,f_2\left(x,p^2,m_1,m_2\right)\,
\left(1 - \ln{f_2\left(x,p^2,m_1,m_2\right)} - \ln{\mu^2}\right)+
\nnu \\
&\frac{i}{16 \pi^2}\,p\cdot p_1\ p\cdot p_2\;
\left[\int_0^1 dx\, x^2 \, \left( - \ln{f_2\left(x,p^2,m_1,m_2\right)}
+ \ln{\mu^2}\right)\right],
\eea
\vspace{-5mm}
\bea
& \int  \frac{d^d q}{\left(2 \pi\right)^d} \,
\frac{1}{\left[ (q + p)^2 - m_1^2 \right] ( q^2 - m_2^2) ( q^2 - m_3^2 ) }
=\hspace{9cm}\nnu\\
& \frac{-i}{16 \pi^2} \int_0^1 dx \ \int_0^{1-x} dy
f_3^{-1}\left(x,y,p^2,m_1,m_2,m_3\right),  \nnu
\eea
\vspace{-5mm}
\bea
& \int  \frac{d^d q}{\left(2 \pi\right)^d} \,
\frac{(q \cdot p_1 )}{\left[ (q + p)^2 - m_1^2 \right]
( q^2 - m_2^2) ( q^2 - m_3^2 ) }=\hspace{9cm}\nnu\\
& \frac{i}{16 \pi^2}\ p\cdot p_1\ \int_0^1 dx \ \int_0^{1-x} dy\ x
f_3^{-1}\left(x,y,p^2,m_1,m_2,m_3\right), \nnu
\eea
\vspace{-5mm}
\bea
&\int  \frac{d^d q}{\left(2 \pi\right)^d} \,  \frac{(q \cdot p_1 ) (q \cdot p_2 ) }
{\left[ (q + p)^2 - m_1^2 \right] ( q^2 - m_2^2) ( q^2 - m_3^2 )
}=\hspace{9cm}\nnu\\
& \frac{-i}{16 \pi^2}\,p\cdot p_1\ p\cdot p_2\;
\int_0^1 dx \
\int_0^{1-x} dy\ x^2 \ f_3^{-1}\left(x,y,p^2,m_1,m_2,m_3\right)
 \nnu \\
& +\frac{i}{32 \pi^2}\, p_1\cdot p_2\;
\left[- \int_0^1 dx \
\int_0^{1-x} dy \ \ln{f_3\left(x,y,p^2,m_1,m_2,m_3\right)} +
\ln{\mu^2}\right].
\nnu
\eea
}
It should be understood that the above result for loop integrals are used after expanding them
properly about the origin of the Diltaz plot variables $(X=Y=0)$ up to quadratic terms in $X, Y$.

The integral formulas needed to compute analytically  the imaginary part of
the Feynman diagrams are:
{\footnotesize
\bea
&\left(2 \pi i \right)^2
\int\ \frac{d^4 q}{\left(2 \pi\right)^4} \ \delta \left(q^2 -
m_\pi^2\right)\ \delta \left[(q+p)^2 -m_\pi^2\right]=\hspace{9cm}\nnu\\
&- \frac{1}{16 \pi}\ \sqrt{\left(1 - \frac{4 m_\pi^2}{p^2}\right)},\nnu
\eea
\bea
&\left(2 \pi i \right)^2 \int\  \frac{d^4 q}{\left(2 \pi\right)^4}
\ q\cdot p_1\ \delta \left(q^2 - m_\pi^2\right) \delta \left[(q+p)^2 -m_\pi^2\right]
=\hspace{6cm}\nnu\\
&\frac{1}{32 \pi}\ \sqrt{\left(1 - \frac{4 m_\pi^2}{p^2}\right)}\;\;\; p\cdot p_1,
\eea
\bea
&\left(2 \pi i \right)^2 \int\  \frac{d^4 q}{\left(2 \pi\right)^4}
\ q\cdot p_1 \,\, q \cdot p_2 \
\delta \left(q^2 - m_\pi^2\right)\ \delta \left[(q+p)^2 -m_\pi^2\right]
=\hspace{9cm}\nnu \\
 & \frac{1}{192 \pi}
 \left(1 - \frac{4 m_\pi^2}{p^2}\right)^{\frac{3}{2}}\;\;\; p^2\ p_1\cdot p_2
-\frac{1}{48 \pi} \left(1 -
\frac{m_\pi^2}{p^2}\right)
\sqrt{\left(1 - \frac{4 m_\pi^2}{p^2}\right)}\;\;\; p\cdot p_1\ p\cdot p_2, \nnu \\
\eea
}
where $\delta$ denotes the Dirac delta function.

The weak chiral Lagrangian according to the convention followed in
Ref.\cite{wise} is
\bea
{\cal L}^{(2)}_{\Delta S = 1} & = & g_{\underline{8}}  \Tr \left(
\lambda^3_2 D_\mu \Sigma^{\dag} D^\mu \Sigma\right) \ \nnu \\
&   & +
g_{\underline{27}} \left[
\Tr \left( \lambda^1_2 \Sigma^{\dag} D_\mu \Sigma \right)
\Tr \left( \lambda^3_1 \Sigma^{\dag} D^\mu  \Sigma \right) +
\Tr \left( \lambda^3_2 \Sigma^{\dag} D_\mu\Sigma \right)
\Tr \left(  \lambda^1_1 \Sigma^{\dag} D^\mu \Sigma \right) \right.  \nnu \\
&  & - \left.
\Tr \left( \lambda^3_2 \Sigma^{\dag} D_\mu \Sigma \right)
\Tr \left( \lambda^2_2 \Sigma^{\dag} D^\mu  \Sigma \right)\right].  \nnu \\
\label{chiwise}
\eea
The $SU(3)$ projection for the twenty-seven part are \cite{swart}
\bea
|\underline{27},\frac{1}{2}\rangle & = & \Tr \left( \lambda^1_2 \Sigma^{\dag} D_\mu \Sigma \right)
\Tr \left( \lambda^3_1 \Sigma^{\dag} D^\mu  \Sigma \right) +
4\ \Tr \left( \lambda^3_2 \Sigma^{\dag} D_\mu\Sigma \right)
\Tr \left(  \lambda^1_1 \Sigma^{\dag} D^\mu \Sigma \right)   \nnu \\
&  & + 5
\Tr \left( \lambda^3_2 \Sigma^{\dag} D_\mu \Sigma \right)
\Tr \left( \lambda^2_2 \Sigma^{\dag} D^\mu  \Sigma \right),  \nnu \\
|\underline{27},\frac{3}{2}\rangle & = & \Tr \left( \lambda^1_2 \Sigma^{\dag} D_\mu \Sigma \right)
\Tr \left( \lambda^3_1 \Sigma^{\dag} D^\mu  \Sigma \right) +
 \Tr \left( \lambda^3_2 \Sigma^{\dag} D_\mu\Sigma \right)
\Tr \left(  \lambda^1_1 \Sigma^{\dag} D^\mu \Sigma \right)   \nnu \\
&  & -
\Tr \left( \lambda^3_2 \Sigma^{\dag} D_\mu \Sigma \right)
\Tr \left( \lambda^2_2 \Sigma^{\dag} D^\mu  \Sigma \right).  \nnu \\
\label{comp}
\eea
Therefore, we have
\bea
|\underline{27}\rangle & = & \frac{5}{9} |\underline{27},\frac{3}{2}\rangle +
\frac{1}{9} |\underline{27},\frac{1}{2}\rangle, \nnu \\
&=&
\frac{2}{3} \Tr \left( \lambda^1_2 \Sigma^{\dag} D_\mu \Sigma \right)
\Tr \left( \lambda^3_1 \Sigma^{\dag} D^\mu  \Sigma \right) +
\Tr \left( \lambda^3_2 \Sigma^{\dag} D_\mu\Sigma \right)
\Tr \left(  \lambda^1_1 \Sigma^{\dag} D^\mu \Sigma \right).
\eea
So, only the $|\underline{27},\frac{3}{2}\rangle$ component was
considered in Ref.~\refcite{wise}.
This component induces the $\Delta I = 3/2$ transition.

The full imaginary part of the amplitude $K^+ \rightarrow \pi^+ \pi^+ \pi^-$ for the octet part
of the weak chiral Lagrangian given by \eq{chiwise} is

{\footnotesize
 \bea
\Im\left[A^{(8)}\left(K^+\rightarrow \pi^+ \pi^+ \pi^-\right)\right]= \hspace{9cm} \nnu \\
  \frac{-1}{192\ \pi\ f_\pi^6}\ \left[  6\ \sqrt{1 - \frac{4\ m_\pi^2}{s_3}} \left(m_K^2 + m_\pi^2
- s_1 - s_2 \right) \left(2\ m_\pi^2 - s_1 - s_2\right) \right.   \nnu \\
   +  2\ \sqrt{1 - \frac{4\ m_\pi^2}{s_1}} \left(9\ m_\pi^4 + m_K^2\ \left(2\ m_\pi^2 + s_1\right)
 + s_1\ \left(5\ s_1 + s_2\right) - m_\pi^2\ \left(11\ s_1 + 4\ s_2\right)\right) \nnu \\
  + \left.  2\ \sqrt{1 - \frac{4\ m_\pi^2}{s_2}} \left(9\ m_\pi^4 + m_K^2 \left(2\ m_\pi^2 + s_2\right)
 + s_2\ \left(5\ s_2 + s_1\right) - m_\pi^2\ \left(11\ s_2 + 4\ s_1\right)\right)
\right], \nnu \\
\label{comon8}
\eea }
and for the twenty-seven part
{\footnotesize
 \bea
\Im\left[A^{(27)}\left(K^+\rightarrow \pi^+ \pi^+ \pi^-\right)\right]= \hspace{9cm} \nnu \\
 \frac{-1}{192\ \pi\ f_\pi^6}\ \left\{ 12\ \sqrt{1 - \frac{4\ m_\pi^2}{s_3}}
\left(m_K^2 + m_\pi^2 - s_1 - s_2\right)\,\left(3\ m_k^2 + 8\ m_\pi^2 - 4\ \left(s_1 + s_2\right)\right)
\right. \nnu \\
+\ \frac{1}{m_K^2 - m_\pi^2} \sqrt{1 - \frac{4\ m_\pi^2}{s_1}}\left[
m_k^4\ \left(23\ m_\pi^2 - 11\ s_1\right)\right. \hspace{3cm} \nnu \\
+  m_k^2\left(31\ m_\pi^4 + 2\ m_\pi^2\left(3\ s_1 - 32\ s_2\right) -
 s_1 \left(s_1 - 16\ s_2\right)
\right)
\nnu \\
\left. - 2\ m_\pi^2 \left(45\ m_\pi^4 + s_1\ \left(13 s_1 + 8\ s_2\right)
- 2\ m_\pi^2 \left(17\ s_1 + 16\ s_2 \right)\right) \right] \nnu \\
+\ \frac{1}{m_K^2 - m_\pi^2} \sqrt{1 - \frac{4\ m_\pi^2}{s_2}}\left[
m_k^4\ \left(23\ m_\pi^2 - 11\ s_2\right)\right. \hspace{3cm} \nnu \\
+  m_k^2\left(31\ m_\pi^4 + 2\ m_\pi^2\left(3\ s_2 - 32\ s_1\right) -
 s_2 \left(s_2 - 16\ s_1\right)\right)
\nnu \\
\left. \left. - 2\ m_\pi^2 \left(45\ m_\pi^4 + s_2\ \left(13 s_2 + 8\ s_1\right)
- 2\ m_\pi^2 \left(17\ s_2 + 16\ s_1 \right)\right) \right]\right\}. \nnu \\
\label{ourc23}
\eea
}
The authors of Ref.\cite{wise} did neglect
terms of $O(m_\pi^2)$ in the weak vertex. But their result for the
octet parts agrees with ours. For the twenty-seven part, there is
disagreements in terms proportional $m_\pi^2$, which can be safely
neglected.
Here is the result found for the twenty seven part.~\cite{wise}
{\footnotesize
 \bea
\Im\left[A^{(27)}\left(K^+\rightarrow \pi^+ \pi^+ \pi^-\right)\right]= \hspace{9cm} \nnu \\
 \frac{-1}{192\ \pi\ f_\pi^6}\ \left\{ 12\ \sqrt{1 - \frac{4\ m_\pi^2}{s_3}}
\left(m_K^2 + m_\pi^2 - s_1 - s_2\right)\,\left(3\ m_k^2 + 8\ m_\pi^2 - 4\ \left(s_1 + s_2\right)\right)
\right. \nnu \\
-\  \sqrt{1 - \frac{4\ m_\pi^2}{s_1}}
\left(-63\ m_\pi^4 + s_2\ (-16\ s_1 + s_2) +
          2\ m_\pi^2\ (32\ s_1 + 7\ s_2) + m_k^2\ (-23\ m_\pi^2 + 11\ s_2)\right)  \nnu \\
-\ \left.  \sqrt{1 - \frac{4\ m_\pi^2}{s_2}}
\left(-63\ m_\pi^4 + s_1\ (-16\ s_2 + s_1) +
          2\ m_\pi^2\ (32\ s_2 + 7\ s_1) + m_k^2\ (-23\ m_\pi^2 + 11\ s_1)\right)
          \right\}. \nnu \\
\label{wisec23}
 \eea
}

Some identities for the imaginary part could be derived, but first
let us simplify our notations as  follows \bea \Tr \left(
\lambda^1_2 \Sigma^{\dag} D_\mu \Sigma \right) \Tr \left(
\lambda^3_1 \Sigma^{\dag} D^\mu  \Sigma \right) \rightarrow G_a,
\nnu \\
\Tr \left( \lambda^3_2 \Sigma^{\dag} D_\mu\Sigma \right)
\Tr \left(  \lambda^1_1 \Sigma^{\dag} D^\mu \Sigma \right) \rightarrow
G_b,
\nnu \\
\Tr \left( \lambda^3_2 \Sigma^{\dag} D_\mu \Sigma \right)
\Tr \left( \lambda^2_2 \Sigma^{\dag} D^\mu  \Sigma \right) \rightarrow
G_{32}.
\nnu \\
\eea
Following the convention of Ref.~\refcite{meson} for the twenty-seven
\bea
|\underline{27}\rangle & = & \frac{5}{3} |\underline{27},\frac{3}{2}\rangle +
\frac{1}{3} |\underline{27},\frac{1}{2}\rangle . \nnu \\
\eea
We verify that $G_{32}$ has the same Feynman rules as $G_b$
but with opposite sign in the case of vertices containing one Kaon
and three pions. Thus, it follows that
$|\underline{27},\frac{1}{2}\rangle$ as defined in \eq{comp}
behaves  as an octet (keeping in mind that the term $G_b - G_a$
behaves as an octet). Consequently the contribution of the
$|\underline{27}\rangle$ must be one-third of that of the octet
for $\Delta I = 1/2$ isospin amplitudes. This is only applicable
for the imaginary part. In fact, our result for the imaginary
parts in Table~(\ref{complex}) satisfies these identities, but it is not
for the results produced in Ref.~\refcite{meson}.

%

\end{document}